\documentclass{aastex631}

\submitjournal{AJ}

\shorttitle{Double line binaries}
\shortauthors{Anguita et al.}

\graphicspath{{./}{figures/}}

\begin{document}


\title{Orbital elements and individual component masses from joint spectroscopic and astrometric data of double-line spectroscopic binaries\footnote{Based in part on observations obtained at the international Gemini Observatory, a program of NSF’s NOIRLab, which is managed by the Association of Universities for Research in Astronomy (AURA) under a cooperative agreement with the National Science Foundation on behalf of the Gemini Observatory partnership: the National Science Foundation (United States), National Research Council (Canada), Agencia Nacional de Investigaci\'on y Desarrollo (Chile), Ministerio de Ciencia, Tecnolog\'{i}a e Innovaci\'on (Argentina), Minist\'erio da Ci\^{e}ncia, Tecnologia, Inova\c{c}\~{a}es e Comunica\c{c}\~{o}es (Brazil), and Korea Astronomy and Space Science Institute (Republic of Korea). Based also in part on observations obtained at the Southern Astrophysical Research (SOAR) telescope, which is a joint project of the Minist\'{e}rio da Ci\^{e}ncia, Tecnologia, e Inova\c{c}\~{a}oes (MCTI) da Rep\'{u}blica Federativa do Brasil, the U.S. National Optical Astronomy Observatory (NOAO), the University of North Carolina at Chapel Hill (UNC), and Michigan State University (MSU).}}

\correspondingauthor{Rene A. Mendez}
\email{rmendez@uchile.cl}

\author{Jennifer Anguita-Aguero}
\affiliation{Astronomy Department \\ Universidad de Chile \\ Casilla 36-D, Santiago, Chile}

\author[0000-0003-1454-0596]{Rene A. Mendez}
\affiliation{Astronomy Department \\ Universidad de Chile \\ Casilla 36-D, Santiago, Chile}
  
\author{Rub\'en M. Claver\'ia}
\affiliation{Department of Engineering \\ University of Cambridge, UK}

\author{Edgardo Costa}
\affiliation{Astronomy Department \\ Universidad de Chile \\ Casilla 36-D, Santiago, Chile}

\begin{abstract}
We present orbital elements, orbital parallaxes and individual component masses, for fourteen spatially resolved double-line spectroscopic binaries derived doing a simultaneous fit of their visual orbit and radial velocity curve. This was done by means a Markov Chain Monte Carlo code developed by our group, which produces posterior distribution functions and error estimates for all the parameters. Of this sample, six systems had high quality previous studies and were included as benchmarks to test our procedures, but even in these cases we could improve the previous orbits by adding recent data from our survey of southern binaries being carried out with the HRCam and ZORRO speckle cameras at the SOAR 4.1m and Gemini South 8.1m telescopes, respectively. We also give results for eight objects that did not have a published combined orbital solution, one of which did not have a visual orbit either.
We could determine mass ratios with a typical uncertainty of less than 1\%, mass sums with uncertainties of about 1\% and individual component masses with a formal uncertainty of $0.01 M_\odot$ in the best cases. A comparison of our orbital parallaxes with available trigonometric parallaxes from Hipparcos and Gaia eDR3, shows a good correspondence; the mean value of the differences being consistent with zero within the errors of both catalogs. 
We also present observational HR diagrams for our sample of binaries, which in combination with isochrones from different sources allowed us to asses their evolutionary status of and also the quality of their photometry.
\end{abstract}


\section{Introduction} \label{sec:intro}

Binary stars are powerful laboratories to test various aspects of stellar astrophysics, because they enable to us to access a key but elusive parameter which dictates the structure and evolution of stars: their mass. In combination with Kepler's laws, the observation of binary systems allows to determine directly the masses of stars. Indeed, the most fundamental parameter determining the internal structure and evolutionary path of stars of a given chemical composition is their initial mass as shown by the well-known Vogt-Russell theorem (\citet{kah72,Kippet12}, for a general review see \citet{Masset01}; and the textbooks by \citet{Iben12} for details of the physical models). This theoretical prediction is however not straightforward to test due to the limited number of stars with well-known individual masses.

Kepler's laws give us the way to determine directly the mass of a stellar system by studying the motion of stars that are bound by their mutual gravitational attraction, i.e., binary stars \citep{pourbaix1994trial}. Considering that roughly half of solar-type stars in the Solar Neighborhood belong to binary systems \citep{Raget10,Duchet13, Fuhret17}, in principle it is possible to determine precise masses for a very large number of stars. Gravitational microlensing might eventually become another potentially very precise method for mass determination \citep{Ghoshet04,Gould14}, albeit so far it has been restricted to a few cases \citep{Bennet20, WyrMan2020}. Circumstelar disks around young stars \citep{Pegueset2021} is also becoming a viable and promising method. In the case of microlensing events, the mass of the lens can be determined only in limited cases, because it requires a knowledge of both the source and lens distances, as well as their relative proper motions. The second method relies on the existence of a purely Keplerian disk\footnote{Usually found only on young stars.}, i.e., in a steady-state configuration, and not subject to magneto-hydrodynamical effects, which enables a purely dynamical mass determination. In the case of visual binary stars, the subject of this paper, a mass determination requires a determination of the so-called orbital elements that completely define the projected orbit in terms of the true intrinsic orbital parameters.

One of the main relationships reflecting the dependency on mass of the star's properties is the mass-luminosity relation (MLR), first discovered empirically in the early 20th century, and later explained on theoretical grounds \citep{Edd24}. Improving the observational MLR is not a simple task, because it involves not only determining precise masses, but also another elusive parameter: distance, by means of trigonometric parallaxes. To complicate things further, the observational MLR has a statistical dispersion which cannot be explained exclusively by observational errors in the luminosity or mass; there seems to be an intrinsic dispersion caused by differences in age and/or chemical composition from star to star. Currently, the best MLR for main sequence stars are those of \citet{Torreset10} and \citet{Beneet16}, but neither of them include low metallicity stars (e.g., only one with $[Fe/H] < -0.25$ in \citet{Torreset10}). Other studies, using long-baseline optical interferometry of binary systems, have begun to address metallicity (e.g. \citet{Bojaet12a, Bojaet12b, FeiChab12}, but have reached only as low as about $[Fe/H] = -0.5$. For a recent study of the effects of metallicity on the MLR for $M < 0.7 M_\odot$ see \citet{Mannet19}. Our own speckle survey, described below, is focusing, in part, precisely on low-metallicity objects, following the pioneer study of \citet{Horchet2015b,Horchet2019}.

The recent advent of the Gaia satellite \citep{Luriet18} has dramatically improved the precision of stellar distances within the Solar Neighborhood. Up to a distance of 250 pc, well beyond what it is usually adopted as the “radius” of the Solar Neighborhood, a parallax determined by Gaia has an uncertainty under 1\%, which by current standards would have basically solved the distance dilemma in the MLR. Despite this promising scenario, much remains to be done to increase the number of stars with well known masses, because of the relative lack of high precision orbits for binary systems. On the other hand, the Gaia satellite faces observational difficulties at resolving systems close to its angular resolution limit. It is well known that Hipparcos parallaxes were indeed biased due to the orbital motion of the binary (i.e., the parallax and orbit signal are blended), as shown by \cite{Soder1999} (see, in particular his Section~3.1, and Table~2), and it is likely that Gaia will suffer from a similar problem\footnote{For example, according to Tokovinin's multiple star catalogue, HIP 64421 contains a binary with a 27~yr orbit. Its Hipparcos parallax is 8.6~mas, its dynamical parallax is 8.44~mas, and its Gaia DR2 parallax is 3~mas. However, Gaia does give a consistent parallax for the C component at 1.9~arcsec: $9.7 \pm 0.3$~mas, see \url{http://www.ctio.noao.edu/~atokovin/stars/}. There are other examples like this in the cited catalog.}.

A good starting point for systematic surveys to determine stellar masses are all-sky catalogues which include identification of confirmed or suspected visual binaries, such as the Hipparcos catalogue \citep{Liet97}, as well as the more recent Gaia discoveries \citep{Keret19, ElBadryet21, Brandt21}; or spectroscopic binaries, such as the Geneva-Copenhagen spectroscopic survey \citep{Noet04}. To this end, in 2014 we initiated  a systematic campaign to complete or improve the observation of southern binaries from mainly the above catalogues \citep{Mendet2018}, using the high-speed speckle camera HRCAM at the SOAR 4.1m telescope \citep{TokCan2008, Toko2018}; several publications have resulted from this effort, \citet{Docoet16, Mendezetal2017, Clavet2019, Docoet19, Mendet2021, Villeet2021, Goet21}. Considering that metal-poor binary systems are typically farther away and therefore fainter and/or more compact spatially, making them difficult objects for optical interferometry with 4m or smaller telescopes, in 2019 we also started a program focused on very these systems with the ZORRO Speckle Camera of the Gemini South (GS) 8.1 m telescope at Cerro Pachón\footnote{See \url{https://www.gemini.edu/instrumentation/current-instruments/alopeke-zorro}}. We note that southern binaries are currently being monitored systematically only by our team, which makes our speckle survey unique. Complementarity with Gaia is a very strong reason to carry out a survey of nearby binaries now: During each observation Gaia is not expected to resolve systems closer than about 0.4~arcsec, though over the mission there will be a final resolution of 0.1~arcsec. This is shown graphically in Figure~1 from \citet{Ziegleret2018}, where the current resolution of the second Gaia data release is shown to be around 1~arcsec, being a function of the magnitude difference between primary and secondary\footnote{It is expected that, from the third Gaia data release and on, the treatment of binary stars will be much improved, by incorporating orbital motion (and its impact on the photocenter position of unresolved pairs) into the overall astrometric solution, thus suppressing/alleviating the parallax bias significantly, this in turns calls precisely for having good orbital elements for these binaries, which is one of the secondary goals of our project.}

For many years, exoplanet searches excluded binary systems but, nowadays, more than 200 planets have been discovered in them\footnote{As of November 3rd 2021,  217 planets are known in 154 binary systems (https://www.univie.ac.at/adg/schwarz/multiple.html).}, representing a multiplicity rate of about 23\% for hosts to exoplanets across all spectral types \citealt{2021FrASS...8...16F}. Initially they were discovered serendipitously, like Gliese 86\,Ab \citep{2000A&A...354...99Q} and $\gamma$ Cep\,Ab \citep{2003ApJ...599.1383H} but, more recently, as part of dedicated imaging, transit and radial velocity (RV) surveys (e.g., see \citet{2021FrASS...8...16F}). Given that the formation of stars in multiple systems is a frequent byproduct of stellar formation, a current open question is to understand how the presence of a stellar companion can affect the planetary formation process. For years, planetary formation theories have been restricted to the case of a single star environment to understand the formation of our own Solar System \citep{2008ASPC..398..235M}. For the most frequent dynamical configuration observed for planet(s) in binaries (see classification by \citealt{1982OAWMN.191..423D}), the S-type circumprimary one with a planet orbiting one component of the binary, generally the most massive one, models predict that the presence of a very close binary companion can truncate a protoplanetary disk, hence obstructing the formation of a planet by core accretion, or ejecting the planet in unstable systems \citep{2015pes..book..309T}. Dedicated observing campaigns confirmed that short and intermediate-separation ($\leq 300$\,au) binaries have statistically less chance to host planets or brown dwarf companions, and that wide binaries on the contrary have no influence on the architectures of planetary systems \citep{2018AJ....156...31M}. This effect seems to be also corroborated by the study of young stars for which short-separation ($\leq 100$\,au) binaries have a lower probably of hosting circumstellar dust in the innermost few AU around each star, therefore with a depleted reservoir of solids for the formation of planets by core accretion \citep{2010ApJ...709L.114D}.

Among different types of binary systems, spatially resolved double-line spectroscopic binaries (hereafter SB2) with known RV curves are particularly important. If their RV curves can be combined with their astrometric orbits, it is possible to obtain a complete and unambiguous solution for the orbital elements, as well as individual component masses with high precision {\citep{Mendezetal2017, Mendet2021}. With high-quality data in hand, it is also possible to derive parallax-free distances -the so-called orbital parallaxes- for these systems, which are derived from the ratio of the semi-major axes \citep{Docoet2018, Picc2020}. Orbital parallaxes are completely independent of the trigonometric parallax, and thus allow an assessment of Gaia’s parallaxes \citep{Pour2000, Mas2015}. Furthermore, increasing the sample of well-studied SB2 is important because, statistically, the mass-ratio distribution of these binaries - a parameter assumed to be frozen since their formation, but observationally retrievable - has an important imprint from the initial mass function, as shown by, e.g., the simulations by \citet{Ducatiet2011}. This traditional vision is however being somewhat challenged by more recent simulations which demonstrate that high-mass stars can capture lower mass stars with luminosities far smaller than those of their host during the first few million years of star cluster evolution \citep{Wallet19}. Additionally, N-body simulations that incorporate magnetohydrodynamic effects have shown that dynamical interactions between stars in the presence of gas during cluster formation can modify the initial mass-ratio distribution towards binaries with larger mass difference \citep{CoClet21}.

Unfortunately, and despite sustained efforts to monitor SB2s (see e.g. \citet{HalKie2020} and their series of papers), the number of those systems for which both a RV curve and a precise astrometric orbit is available is still rather small. In this paper, we contribute to alleviate this situation by determining combined orbits for 14 SB2 systems, some of which have been observed during our SOAR speckle survey of southern binary stars \citep{Mendet2018} and for which there is complementary published data as well.

Methodologically, the most common procedure is to solve the astrometric visual orbit separately from the RV curve. In this approach, the amplitude ratio between the RV curves of the primary and secondary gives an estimation of the mass ratio, while the astrometric orbit gives the mass sum (assuming a parallax). This is most often done when the primary and secondary amplitudes are not well determined, e.g., when the spectral resolution prevents full de-blending of the spectral lines of both components (see e.g., \citet{TokLat2017}), or in the case of single-line spectroscopic binaries with a visual orbit (see, e.g., \citet{Docoet2018singleline}). In both cases, it is not possible to directly link the RV curve to the astrometric orbit in a self-consistent manner. On the other hand, if the visual orbit and the velocity amplitudes are believed to be reliable, it is possible to determine a simultaneous solution, and in the process determine the orbital parallax based solely on the orbital motion of the pair. Both scenarios are thoroughly explained, including graphical flowcharts, in \citet{Villeet2021}, Section~4.2 of that paper, while in Appendix~A of \citet{Mendezetal2017} we provide a detailed step-by-step flow of our calculations}. In this work, well measured RV curves and visual orbits are available for most systems, so by adopting the latter of these schemes we were able to determine orbital parallaxes for the majority of our binaries.

This paper is organized as follows: In Section~\ref{sec:sample} we introduce our list of SB2s together with their basic photometric properties. In Section~\ref{sec:orbs} we present the results of our orbital calculations and individual component masses. These results, along with the photometry, are presented on HR diagrams in Section~\ref{sec:hrdiag}, while in Section~\ref{sec:objnotes} we describe our sample on an object-by-object basis. Finally, in Section~\ref{sec:concl} we present the main conclusions of our study.

\section{Basic properties of the Binary Systems}\label{sec:sample}

To select the sample for the present work, we started by doing a cross-match between the Sixth Catalog of Orbits of Visual Binary Stars maintained by the US Naval Observatory (hereafter Orb6\footnote{Available at \url{https://www.usno.navy.mil/USNO/astrometry/optical-IR-prod/wds/orb6}}) and the
9th Catalog of Spectroscopic Binary Orbits (hereafter SB9\footnote{Updated regularly, and available at \url{https://sb9.astro.ulb.ac.be/}}, \citet{Pouret2004}). Orb6 is the most comprehensive catalog of binary systems with published orbital elements, and SB9 contains RV amplitudes for all binary systems for which it has been possible to fit a RV curve. Having identified those systems confirmed as double-line spectroscopic binaries in SB9, we pinpointed the binaries for which a combined astrometric/RV study of the orbit was not available in the literature by means of the notes and comments given in Orb6 and SB9. This lead to an initial working list of 17 binary systems.


For the systems selected as indicated above, we retrieved their RV data from SB9 or from references provided therein, astrometric data from the US Naval Observatory Fourth Catalog of Interferometric Measurements of Binary Stars\footnote{The latest version, called int4, is available at \url{https://www.usno.navy.mil/USNO/astrometry/optical-IR-prod/wds/int4/fourth-catalog-of-interferometric-measurements-of-binary-stars}} and historical astrometry included in the Washington Double Star Catalog effort (\citet{WDSCat2001}, hereafter WDS, kindly provided to us by Dr. Brian Mason from the US Naval Observatory). Finally, we included recent results obtained with the HRCam speckle camera at the SOAR 4.1m telescope as part of our monitoring of Southern binaries described in \citet{Mendezetal2017}. Examination of the information collected showed that only 14 of the systems in our starting list had sufficient data to warrant further analysis. We must emphasize that, as a result of our selection process, our final sample is very heterogeneous and it is not complete or representative in any astrophysical sense. From this point of view, the main contribution of this paper is the addition of new orbits and individual component masses for this type of binaries. In our final list we also included a few previously studied objects in order to test our procedures and compare results. When available, we added SOAR+HRCam astrometric measurements we secured after the publication of their last orbit determination to improve the solution.

Table~\ref{tab:photom1} presents basic properties available in the literature for the sample studied in this paper. The first three columns give the name in the WDS Catalog (below it, its corresponding HD number), the discoverer designation code assigned in the WDS to each target used throughout the paper, and the sequential number in the Hipparcos catalog. The fourth column gives the apparent $V$ magnitude for the system as listed in SIMBAD ($V_{\mbox{Simbad}}$, \citet{SIMBAD}). The fifth and sixth columns present the $V$ magnitude given on the Hipparcos catalog ($V_{\mbox{Hip}}$) and its source, respectively. The seventh and eighth columns list the color ($(V-I)_{\mbox{Hip}}$) and its source, respectively, also from the Hipparcos catalog. The ninth and tenth columns give the $V$ magnitudes for the primary ($V_{\mbox{A}}$) and secondary ($V_{\mbox{B}}$) components, respectively, as listed in the WDS catalog. As a sanity test, the integrated apparent magnitude for the system $V_{\mbox{t}}$ (from the WDS individual component photometry) is given in the eleventh column\footnote{Computed as $V_{\mbox{t}} = -2.5 \times \log \left( 10^{-0.4 \cdot V_{\mbox{A}}} + 10^{-0.4 \cdot V_{\mbox{B}} } \right)$}. In the twelfth and thirteenth columns we report our own measured magnitude differences in the Str\"omgren y filter ($\Delta y$) and in the Cousins I filter $\Delta I$ ($\equiv I_{\mbox{B}} -I_{\mbox{A}}$ between secondary and primary), respectively. These measurements are part of our speckle binary program mentioned in the previous paragraph. Finally, in the last column we report the spectral type and luminosity class for the primary and secondary, after the $+$ sign, when available, from WDS and SIMBAD, respectively.

\floattable
\rotate
\begin{deluxetable}{cccccccccccccc}
\tablecaption{Object identification and basic photometry.\label{tab:photom1}}
\tabletypesize{\scriptsize}
\tablecolumns{14}
\tablewidth{0pt}
\tablehead{
\colhead{WDS name} &
\colhead{Discoverer} &
\colhead{HIP} & 
\colhead{$V_{\mbox{Simbad}}$\tablenotemark{1}} &
\colhead{$V_{\mbox{Hip}}$\tablenotemark{2}} &
\colhead{Source$_{V_{\mbox{Hip}}}$\tablenotemark{3}} &
\colhead{$(V-I)_{\mbox{Hip}}$\tablenotemark{2}} &
\colhead{Source$_{(V-I)_{\mbox{Hip}}}$\tablenotemark{4}} &
\colhead{$V_{\mbox{A}}$\tablenotemark{5}} &
\colhead{$V_{\mbox{B}}$\tablenotemark{5}} &
\colhead{$V_{\mbox{t}}$} &
\colhead{$\Delta y$\tablenotemark{6}} &
\colhead{$\Delta I$\tablenotemark{6}} &
\colhead{Sp Type} \\
\colhead{HD number} &
\colhead{designation} &
\colhead{number} & 
\colhead{} &
\colhead{} &
\colhead{} &
\colhead{} &
\colhead{} &
\colhead{} &
\colhead{} &
\colhead{} & 
\colhead{} &
\colhead{} &
\colhead{WDS/Simbad}
}
\startdata
00352$-$0336 & HO212AB & 2762 & $5.201 \pm 0.009$ & 5.2 & G & $0.64 \pm 0.02$ & A & 5.61 & 6.9 & 5.32 & $1.33 \pm 0.06$ & $1.17 \pm 0.06$ & F8V/F7V$+$G4V  \\
\textbf{3196} &&&&&&&&&&&&&\\
02128$-$0224 & TOK39Aa,Ab & 10305 & $5.66 \pm 0.01$ & 5.65 & H & $0.63 \pm 0$ & G & 6.18 & 6.69 & 5.65 & $0.35 \pm 0.40$ & --- & F8V/F8V \\
\textbf{13612} &&&&&&&&&&&&&\\
04107$-$0452 & A2801 & 19508 & $7.35 \pm 0.01$ & 7.36 & G & $0.71 \pm 0.00$ & H & 8.3 & 8.3 & 7.55 & $0.80 \pm 0.10$ & $0.57 \pm 0.28$ & G0/G3/5IV \\
\textbf{26441} &&&&&&&&&&&&&\\
04184$+$2135 & MCA14Aa,Ab & 20087 & $5.631 \pm 0.005$ & 5.64 & H & $0.34 \pm 0.03$ & C & 5.6 & 8.1 & 5.50 & --- & --- & F0V/F0V \\
\textbf{27176} &&&&&&&&&&&&&\\
07518$-$1354 & BU101 & 38382 & --- & 5.16 & G & $0.67 \pm 0.02$ & A & 5.61 & 6.49 & 5.21 & $0.93 \pm 0.16$ & $0.83 \pm 0.06$ & G1V/G0V \\
\textbf{64096} &&&&&&&&&&&&&\\
11560$+$3520 & CHR258 & 58184 & $6.74 \pm 0.01$ & 6.74 & H & $0.60 \pm 0.01$ & L & 7.0 & 9.1 & 6.85 & --- & --- & F5/F5 \\
\textbf{103613} &&&&&&&&&&&&&\\
14492$+$1013 & A2983 & 72479 & --- & 8.42 & H & $0.91 \pm 0.01$ & H & 9.27 & 9.36 & 8.56 & $0.22 \pm 0.26$ & $0.43 \pm 0.12$ & K2VK2V \\
\textbf{130669} &&&&&&&&&&&&&\\
15282$-$0921 & BAG25Aa,Ab & 75718 & 6.883 & 6.89 & G & $0.86 \pm 0.02$ & A & 6.9 & 10.2 & 6.85 & $3.52 \pm 0.19$ & $2.28 \pm 0.22$ & G9V/G9V \\
\textbf{137763} &&&&&&&&&&&&&\\
16584$+$3943 & COU1289 & 83064 & $8.09 \pm 0.01$ & 8.09 & H & $0.65 \pm 0.01$ & H & 8.4 & 8.4 & 7.65 & --- & --- & G0/G0 \\
\textbf{153527} &&&&&&&&&&&&&\\
18384$-$0312 & A88AB & 91394 & $6.482  \pm 0.010$ & 6.49 & G & $0.64 \pm 0.07$ & F & 7.22 & 7.51 & 6.60 & $0.15  \pm 0.20$ & 0.0 & F9V/F8V \\
\textbf{172088} &&&&&&&&&&&&&\\
20102$+$4357 & STT400 & 99376 & --- & 7.41 & H & $0.73 \pm 0.00$ & H & 7.6 & 9.83 & 7.47 & --- & --- & G3V/G4V$+$G8V \\
\textbf{191854} &&&&&&&&&&&&&\\
20205$+$4351 & IOT2Aa,Ab & 100287 & 6.85 & 6.78 & G & $0.53 \pm 0.00$ & H & 6.85 & --- & --- & --- & --- & WR+O/WC7p$+$O5 \\
\textbf{193793} &&&&&&&&&&&&&\\
20527$+$4607 & A750 & 103055 & $8.709 \pm 0.015$ & 8.65 & H & $0.83 \pm 0.01$ & L & 9.11 & 10.26 & 8.79 & --- & --- & G8V/G8V \\
\textbf{--} &&&&&&&&&&&&&\\
23485$+$2539 & DSG8 & 117415 & $7.070 \pm 0.009$ & 7.08 & H & $0.51 \pm 0.01$ & L & 7.8 & 7.9 & 7.10 & --- & --- & F5V$+$F5.5V/F2IV$-$V \\
\textbf{223323} &&&&&&&&&&&&&\\
\enddata
\tablenotetext{1}{From the SIMBAD database.
\tablenotetext{2}{From the Hipparcos mission, VizieR catalog I/239.}
\tablenotetext{3}{G=ground-based, H=HIP.}
\tablenotetext{4}{"A" for an observation of $V-I$ in the Cousins system, "F", "G", "H" and "I" when $V-I$ was derived from measurements in other  bands/photoelectric systems; "L" when $V-I$ was derived from Hipparcos and Star Mapper photometry.}
\tablenotetext{5}{From the WDS catalog.}
\tablenotetext{6}{From our own HRCam@SOAR measurements in the Str\"omgren y-band or in the Cousins $I$-band, respectively. When more than one measurement, it is the average.
When more than three measurements, we quote the standard deviation on the mean.}}
\end{deluxetable}

Precise photometry is required to place the individual components in an HR diagram (Section~\ref{sec:concl}). While there is an overall good agreement between SIMBAD, Hipparcos, and the combined ($V_{\mbox{t}}$) magnitudes, the quality of the photometry presented in Table~\ref{tab:photom1} is somewhat variable, as can be readily seen by comparing the fourth, fifth, and eleventh columns of that table. Therefore, in order to increase our comparison basis, we have searched for additional photometry of our targets in more recent all-sky photometric surveys for bright stars; in particular, in the "The All Sky Automated Survey" (ASAS\footnote{\url{http://www.astrouw.edu.pl/asas/?page=main}}
\citet{Pojmanski1997}), the "All-Sky Automated Survey for Supernovae"
(ASAS-SN\footnote{\url{http://www.astronomy.ohio-state.edu/asassn/index.shtml}},
\citet{Kochaneket2017,Jayasingheet19}) and "The AAVSO Photometric All-Sky Survey" (APASS\footnote{\url{https://www.aavso.org/apass}}, \citet{Hendenet2009}, data release 10, November 2018).
All three catalogs report $V$-band magnitudes. APASS includes in addition Sloan $i'$-band photometry, which unfortunately cannot be compared directly with $I$-band values from Hipparcos. To have an extra comparison source in the $I$ bandpass, we used the "All-sky spectrally matched Tycho-2 stars" available at the CDS\footnote{VizieR catalog VI/135}. This catalog presents synthetic photometry in various bands, including $V$ and $I$, from a spectral energy distribution (SED) fit to 2.4 million stars in the Tycho-2 catalog by \citet{PicklesDepagne2010} (PD2010 hereafter). In Table~\ref{tab:photom2} we present the photometry obtained from the above catalogs, together with their quoted uncertainties.
While the above surveys measure and report everything they detect, their
photometry is not reliable at the bright end. Based on the description of the
different surveys, the reliability limit for ASAS, ASAS-SN, and APASS is 8, 10
and 7th mag respectively. Therefore, in Table~\ref{tab:photom2} we indicate
with an asterisk (*) dubious values (up to 1 mag brighter than the bright mag
limit per each survey), and with a double asterisk (**) even brighter objects
whose measurements should be considered as very uncertain. As can be seen from
Table~\ref{tab:photom2}, since most of our targets are quite bright, the
photometry from these surveys is unfortunately not very reliable for some of
them. An important point to make here is that while many of our targets do have high-quality Gaia photometry, it is unfortunately not useful for our purposes due to the special band-passes adopted by the mission \citep{Maiz17}\footnote{See also \url{https://www.cosmos.esa.int/web/gaia/edr3-passbands} and \citet{Rieet21}.}, which do not agree with the band-passes in which we measure magnitude differences at SOAR nor with the passbands used in the WDS catalogue.

\floattable
\begin{deluxetable}{cccccccc}
\tablecaption{Additional photometry from recent all-sky surveys.\label{tab:photom2}}
\tabletypesize{\scriptsize}
\tablecolumns{8}
\tablewidth{0pt}
\tablehead{
\colhead{WDS name} &
\colhead{Discoverer} &
\colhead{$V_{\mbox{ASAS}}$} &
\colhead{$V_{\mbox{ASAS-SN}}$} &
\colhead{$V_{\mbox{APASS}}$} &
\colhead{$i'_{\mbox{APASS}}$} &
\colhead{$V_{\mbox{PD2010}}$} &
\colhead{$I_{\mbox{PD2010}}$} \\
\colhead{HD number} &
\colhead{designation} &
\colhead{} &
\colhead{} &
\colhead{} &
\colhead{} &
\colhead{} &
\colhead{}
}
\startdata
00352$-$0336 & HO212AB & $5.772 \pm 0.306$ \textbf{**} & $5.69 \pm 0.644$ \textbf{**} & $6.157 \pm 0.078$ \textbf{*} & $5.148 \pm 0.012$ & 5.203 & 4.58 \\ 
\textbf{3196} &&&&&&&\\
02128$-$0224 & TOK39Aa,Ab & $6.046 \pm 0.194$ \textbf{**}& --- & $6.084 \pm 0.001$ \textbf{*} & $5.482 \pm 0.001$ & 5.663 & 5.06 \\ 
\textbf{13612} &&&&&&&\\
04107$-$0452 & A2801 & $7.352 \pm 0.34$ \textbf{*} & --- & $7.403 \pm 0.04$ & $7.214 \pm 0.069$ & 7.385 & 6.65 \\
\textbf{26441} &&&&&&&\\
04184$+$2135 & MCA14Aa,Ab & $6.171 \pm 0.319$ \textbf{**} & $6.24 \pm 0.119$  \textbf{**} & --- & --- & 5.625 & 5.29 \\ 
\textbf{27176} &&&&&&&\\
07518$-$1354 & BU101 & $5.91 \pm 0.244$ \textbf{**} & --- & $6.366 \pm 0.049$ \textbf{*} & $5.333 \pm 0.025$ & 5.202 & 4.51 \\
\textbf{64096} &&&&&&&\\
11560$+$3520 & CHR258 & --- & $6.24 \pm 0.119$ \textbf{**} & --- & --- & 6.729 & 6.12 \\ 
\textbf{103613} &&&&&&&\\
14492$+$1013 & A2983 & $8.401 \pm 0.032$ & $8.96 \pm 0.109$ \textbf{**} & $8.403 \pm 0.058$ & $7.772 \pm 0.024$ & 8.421 & 7.54 \\
\textbf{130669} &&&&&&&\\
15282$-$0921 & BAG25Aa,Ab & $6.883 \pm 0.032$ \textbf{**} & $11.39 \pm 1.084$\tablenotemark{1}  & $6.939 \pm 0.111$ \textbf{*} & --- & 6.864 & 5.99 \\
\textbf{137763} &&&&&&&\\
16584$+$3943 & COU1289 & --- & $8.52 \pm 0.094$ \textbf{**} & --- & --- & 8.093 & 7.4 \\
\textbf{153527} &&&&&&&\\
18384$-$0312 & A88AB & $6.543 \pm 0.052$ \textbf{**} & --- & $6.862 \pm 0.01$ \textbf{*} & --- & 6.484 & 5.81 \\
\textbf{172088} &&&&&&&\\
20102$+$4357 & STT400 & --- & $7.92 \pm 0.097$ \textbf{**} & --- & --- & 7.425 & 6.70 \\
\textbf{191854} &&&&&&&\\
20205$+$4351 & IOT2Aa,Ab & --- & $7.67 \pm 0.091$ \textbf{**} & --- & --- & 6.757 & 6.03 \\
\textbf{193793} &&&&&&&\\
20527$+$4607 & A750 & --- & $8.98 \pm 0.09$ \textbf{**} & --- & --- & 8.677 & 7.82 \\
\textbf{--} &&&&&&&\\
23485$+$2539 & DSG8 & $7.057 \pm 0.028$ \textbf{*} & $7.54 \pm 0.114$ \textbf{**} & --- & --- & 7.042 & 6.52 \\
{\bf 223323} &&&&&&&\\
\enddata
\tablenotetext{1}{Five arcsec away from the target, possible miss-identification.}
\end{deluxetable}

\begin{figure}[ht!]
\epsscale{0.9}
\plotone{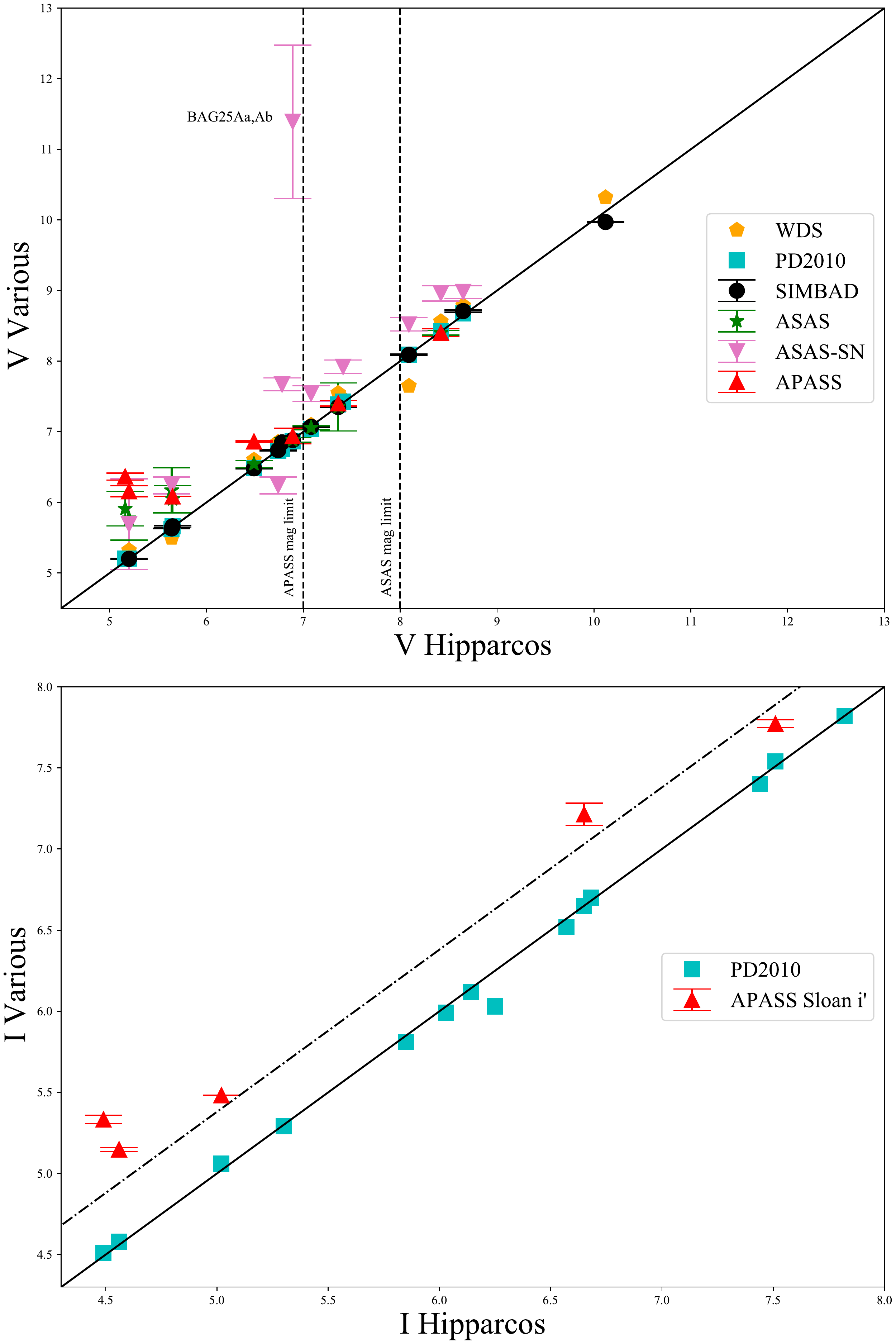}
\caption{Comparison of Hipparcos $V$ (top panel) and $I$-band magnitudes (bottom panel) with other photometric data presented in Tables~\ref{tab:photom1} and~\ref{tab:photom2}. The dotted vertical lines in the upper panel show the reliability bright limits of the APASS ($V=7$) and ASAS ($V=8$) surveys. In the whole magnitude range covered by these figures, {\it all} ASAS-SN photometry is unreliable (note their large declared error bars), but it was included for completeness. In both panels, the diagonal line depicts a one-to-one relationship. In the lower panel, the diagonal dot-dashed line shows the 0.38 mag offset between $I$ and $i'$ from APASS, found by \citet{Mendet2021} (their Figure~1). The highly discrepant point for Bag25Aa,Ab from ASAS-SN in the upper panel is probably due to a miss-identification of the target. See text for details.\label{fig:photom1}}.
\end{figure}

In Figure~\ref{fig:photom1} we show a comparison of the $V$ and $I$ photometry presented in Tables~\ref{tab:photom1} and~\ref{tab:photom2}. We chose to plot Hipparcos magnitudes in the abscissa because it is the largest and most homogeneous data-set for our sample of targets. As can be seen in this figure, there is a relatively good correspondence  in the $V$ band between the photometry from Hipparcos and that from SIMBAD, and also with the combined photometry $V_{\mbox{t}}$ from WDS. The fit of $V_{\mbox{Sim}} {\it vs.} V_{\mbox{Hip}}$ has an rms residual of 0.053~mag, while that of $V_{\mbox{t}}$ {\it vs.} $V_{\mbox{Hip}}$ is 0.17~mag. The larger rms for the latter can be explained mostly by one measurement: $V_{\mbox{t}}$ for COU1289 is too bright in comparison with $V_{\mbox{Hip}}$ (which is the value adopted by SIMBAD too, see Table~\ref{tab:photom1}) - this object is further discussed in Section~\ref{sec:objnotes}. Based on our photometric comparisons, we will thus adopt 0.06~mag as an estimate of the uncertainty of the photometry in Section~\ref{sec:hrdiag}, see also Figures~\ref{fig:hrdiag} and~\ref{fig:hrdiag_2}.

As it can be seen in Figure~\ref{fig:photom1}, at the bright end ($V < 7.0$) ASAS and APASS magnitudes exhibit larger photometric errors and scatter, consistent with their declared bright end reliability. In the case of ASAS-SN this problem extends down to the faintest data plotted with ASAS photometry ($V \sim 9$). ASAS-SN $V$ for object BAG25Aa,Ab is an extreme outlier, but we believe this is due to a miss-identification of another object near the target located at a distance of 5~arscec, because the binary is too bright for the survey (in contrast, the available measurements from the other surveys cluster around the one-to-one relationship). In the range $V > 7.0$ both ASAS and APASS exhibit good consistency, within the errors, between each other, and also with Hipparcos and SIMBAD.

For the $I$ band, the comparison is restricted only to the PD2010 SED fitted photometry \citep{PicklesDepagne2010} and the Sloan $i'$ filter measurements from the APASS survey. As shown in \citet{Mendet2021}, there is an offset of about 0.38~mag between APASS $i'$ and $I$, which is depicted by the dot-dashed line in the plot. After applying this offset, the APASS $i'$-band photometry is commensurable to that derived by Hipparcos and PD2010. In general, we appreciate a good correspondence between Hipparcos, PD2010 and APASS, with an overall rms of the one-to-one fit of 0.06~mag, i.e., similar to the one found for the $V$-band.

\subsection{Individual component magnitudes}\label{sec:comp}

To place our targets in an HR diagram (see Section~\ref{sec:hrdiag}), individual component magnitudes are needed. The $I$-band combined magnitude for each system was computed from $I_{\mbox{Hip}}  = V_{\mbox{Hip}} - (V-I)_{\mbox{Hip}}$, and the individual $I$ component magnitudes as follows: $I_{\mbox{A}} = I_{\mbox{Hip}} + 2.5 \times \log \left( 1.0 + 10^{-0.4 \cdot \Delta I} \right)$ (primary) and $I_{\mbox{B}} = I_{\mbox{A}} + \Delta I$ (secondary). 
For the $V$-band we used the Hipparcos system magnitudes and the WDS magnitude differences - or our own value $\Delta V$ derived from our measured values of $\Delta y$ quoted in Table~\ref{tab:photom1} - depending on the location of the binary on the HR diagram relative to the theoretical isochrones plotted (see Section~\ref{sec:hrdiag}).

From the data in Table~\ref{tab:photom1} it can be seen that the Hipparcos
photometry is in general in good agreement with the photometry for the system
given in WDS ($V_{\mbox{t}}$) and SIMBAD; as well as with that from other
photometric surveys (see Table~\ref{tab:photom2}). Furthermore, if we define $\Delta V = V_{\mbox{B}}-V_{\mbox{A}}$, the mean difference $<\Delta V - \Delta y> = -0.13 \pm 0.32$~mag for seven objects in Table~\ref{tab:photom1} is in agreement with \citet{TokMasHar2010} and \citet{Mendet2021}, indicating that our SOAR magnitude differences seem reliable. This gives us some confidence on the photometry presented, albeit see the extended discussion about this for individual objects on Section~\ref{sec:objnotes}, and Figures~\ref{fig:hrdiag} and~\ref{fig:hrdiag_2} in Section~\ref{sec:hrdiag}.

For one target, IOT2Aa,Ab, individual component magnitudes are not available in WDS, and, being a northern target, it was not observed at SOAR. Several other, mostly northern, targets in Table~\ref{tab:photom1} also lack a measured $\Delta I$, and hence we could not compute individual component magnitudes for them.



\section{Orbital elements, orbital parallaxes, and individual components mass}\label{sec:orbs}

As mentioned in Section~\ref{sec:sample} the astrometric data used in this work is a combination of published information from WDS, with recent measurements made with the HRCam speckle camera\footnote{For up-to-date details of the instrument see \url{http://www.ctio.noao.edu/~atokovin/speckle/}} mounted on the SOAR 4.1 m telescope in the context of the program described in \citet{Mendezetal2017}. Part of our data has not been published yet.\footnote{In the site \url{http://www.das.uchile.cl/~rmendez/B_Research/JAA-RAM-SB2/} we make available our input files, indicating the adopted uncertainty and quadrant flips, if any, for each data entry, and the origin of the measurements in the last column, following the nomenclature in int4.}

Regarding the uncertainty, or equivalent weight, of the historical data that we included in our orbit calculations, we adopted the value indicated in the WDS, when available, or errors typical for the observational procedure used (e.g. interferometric/digital imaging/photographic/micrometer).
On the other hand, HRCam has been shown to deliver a precision of 1-3~mas in angular separation for objects brighter than $V \sim 12$ on a routine basis \citep{Toko2018}. 
In our HRCam@SOAR survey "calibration binaries", binaries with very well known orbits (grades 1 or 2 in Orb6), are observed every night to calibrate our measurements, leading to systematic errors of less than 0.1~deg in position angle, and better than 0.2\% in scale -smaller than our internal precision\footnote{One of these "astrometric standards", WDS07518$-$1354=BU101, is an SB2 and is included in this paper - see Section~\ref{sec:objnotes}}.
The exact final precision of our measurements depends however on a number of factors, but in this paper we will adopt an uncertainty of 3~mas as representative of all our HRCam data. As emphasized in \citet{Mendezetal2017}, one should bear in mind that the assignment of weights to each observational point is somewhat subjective (specially for older data) and  plays a an important role in the orbital solution. Slightly different orbital solutions, from authors using the same astrometric dataset, are in some cases due to different weighting.

Orbits have been derived using our Markov Chain Monte Carlo code (hereafter MCMC). In order to reduce the dimension of the search space, we adopt the parametrization of \citet{Mendezetal2017} in which elements $P$, $T$, $e$, $\Omega$, $i$, mass ratio $q=m_{\text{B}}/m_{\text{A}}$ and parallax $\varpi$ are explored via Markov chain Monte Carlo, whereas $a$ (semi-major axis), $\omega$ and $V_{\text{CoM}}$ are calculated analytically via the exact least-squares solution given ($P$, $T$, $e$, $\Omega$, $i$, $q$,  $\varpi$). Details can be found in \citet{Mendezetal2017}, where an MCMC algorithm is used to carry out joint estimation of orbital parameters and radial velocity (see \citet{Mendezetal2017}, Appendix A for details of the least-squares estimate). 

From a methodological standpoint, the only difference between the algorithm utilised here and that of \citet{Mendezetal2017} is that in the present work we lift the restriction $q = m_{\text{B}}/m_{\text{A}} < 1$. This allows the algorithm to handle uncertainty about the primary and secondary stars: if the mass ratio $q$ is greater than one, the algorithm simply calculates the parameters as if $m_{\text{B}}$ was the primary (this ``swap'' leaves all the parameters other than $\omega$, $\Omega$ untouched). While the value $m_{\text{B}}/m_{\text{A}}$ (i.e., $q$) reported in Table~\ref{tab:plx} is just a natural element of the parametrization adopted (hence a raw output of the MCMC algorithm), amplitudes $K_{\text{A}}$, $K_{\text{B}}$ in Table~\ref{tab:orbel} are calculated as a function of the values of $P$, $e$, $q$, $a$, $\sin i$, $\varpi$ of each MCMC sample.

 Since the code incorporates the parallax of the system as an unknown parameter of the estimation process, it allows us to determine dynamically self-consistent orbital parallaxes, as originally suggested by \citet{Pour2000}. Additionally, SB2s allow a calculation of individual component masses because the astrometric solution gives the mass sum, while the amplitudes of the RV curve gives the mass ratio. Our code produces posterior probability distribution functions (hereafter PDFs) for all the physical and geometrical parameters involved. These PDFs allow us to reliably estimate parameter uncertainties in the following way: It is customary to represent the uncertainties in terms of the dispersion $\sigma$, but this quantity is well defined only for orbits where the PDFs are “well behaved” (e.g., they are symmetrical), and it becomes meaningless for orbits that may exhibit long tails, as in the case of uncertain orbits. For this reason, we instead adopt the upper (third) quartile (Q75) and the lower (first) quartile (Q25) of the distribution as a measure of the spread of the corresponding PDF, and hence as a quantitative measure of the dispersion (uncertainty) of the corresponding parameter\footnote{For a Gaussian function, one can convert
from one to the other using the fact that $\sigma = (Q75- Q25/1.349$.}, which is consistent with the available data and the underlying Keplerian model.

We ran our MCMC routine with a chain length of two~million samples and a burn-in period of $50,000$ iterations. The comparatively short burn-in time is explained on the grounds that the target parameters were initialized favourably: we fed the MCMC routine with approximate initial values from the optimization-based routine ORBIT developed by \citet{Toko1992}\footnote{The code and user manual can be downloaded from \url{http://www.ctio.noao.edu/~atokovin/orbit/index.html}. We note that this code fits the RV amplitudes independently from the visual orbits, and hence it does not compute directly an orbital parallax. For this reason ORBIT was used only to obtain preliminary orbital parameters to initialize our MCMC code.}. From the steadiness of the average value of the orbital parameters over time, we conclude that all the solutions obtained are stable.
It is worth noting that an added benefit of the large number of samples generated is that the resulting PDF histograms look rather smooth (see Figures~\ref{fig:hist} and \ref{fig:hip10305pdf}) despite the fact that they are based directly on the MCMC samples, i.e., they are raw histograms rather than kernel-smoothed densities as in, e.g., \citet{wand1994kernel}.


The results from our MCMC code for the 14 SB2 binaries selected as explained in Section~\ref{sec:sample} are given in Table~\ref{tab:orbel}. In the first two columns we give the WDS name and the source of previous orbital information, if available. In the following columns we present the seven classical orbital elements, the RV for the center-of-mass of the system ($V_{\text{CoM}}$) and the semi-amplitudes for the primary ($K_1$) and the secondary ($K_2$). In the penultimate column we indicate the grade of the orbit according to Orb6 (1: best, 5: worst) and SB9 (5: best, 1: worst), and in the last column the reference to the most recently published astrometric orbit (from Orb6), or RV solution (from SB9).For each object in this table, two sets of values for the orbital elements are provided. The upper row gives the the Maximum Likelihood (hereafter ML) value. For an explanation of why this is the selected estimator, please see the discussion in \citet{Mendet2021} (Section~3.1 on that paper). The lower row gives the median derived from the posterior PDF of the MCMC simulations, as well as the upper quartile ($Q75$) and the lower quartile ($Q25$) of the distribution in the form of a superscript and subscript respectively. As explained before, the quartiles give us an estimation of the uncertainty of our estimated parameters. A look at the results on this Table indicates that our values coincide in general quite well with those from previous studies. In particular, it is well known that the argument of periapsis ($\omega$) is well-determined by RV measurements as long as the distinction between primary and secondary is unambiguous (difficult, e.g., for equal-mass binaries); the table shows our values are indeed quite close to those from SB9, albeit with smaller uncertainties in our case. On the other hand, the longitude of the ascending node ($\Omega$) can be well determined from astrometric observations alone, but it suffers from the same ambiguity in the case of equal-brightness binaries. From the table we see that there is good correspondence between  our values for $\Omega$ and those form Orb6 (but, again, with smaller formal uncertainties in our case), except for five objects. As it will be shown below, four of these objects have values of the mass ratio $q$ quite close to one, which probably explains this discrepancy.

\floattable
\rotate
\begin{deluxetable}{cccccccccccccc}
\tablecaption{Extended orbital elements for our SB2 binaries. \label{tab:orbel}}
\tabletypesize{\tiny}
\tablecolumns{14}
\tablewidth{0pt}
\tablehead{ \colhead{WDS name}&
\colhead{Source} &
\colhead{P} & \colhead{T$_0$} &
\colhead{e} & \colhead{a} & \colhead{$\omega$} & \colhead{$\Omega$} &
\colhead{i} & \colhead{$V_{\text{CoM}}$} & \colhead{$K_1$} & \colhead{$K_2$} & \colhead{Gr} & \colhead{Orbit}\\ \colhead{\textbf{HD number}} &
& \colhead{(yr)} & \colhead{(yr)} &
 & \colhead{(mas)} & \colhead{($^{\circ}$)} & \colhead{($^{\circ}$)} & \colhead{($^{\circ}$)} & \colhead{(km/s)}&\colhead{(km/s)} & \colhead{(km/s)} &\colhead{}&\colhead{Author}
}
\startdata
00352$-$0336 & This paper &$6.8975$ &$1890.6130$ &$0.7616$ &$234.26$ &$104.73$ &$328.34$ &$47.89$ &$9.09$  &$11.67$ &$15.64$ &&\\ \textbf{3196}&
& $6.8975_{-0.0006}^{+0.0005}$ &$1890.6139_{-0.0085}^{+0.0091}$ &$0.7615_{-0.0014}^{+0.0013}$ &$234.07_{-0.73}^{+0.74}$ &$104.70_{-0.18}^{+0.18}$ &$328.39_{-0.22}^{+0.22}$ &$47.83_{-0.24}^{+0.24}$ &$9.08_{-0.03}^{+0.03}$ &$11.68_{-0.51}^{+0.51}$ &$15.59_{-0.17}^{+0.17}$&&\\
& \href{https://www.usno.navy.mil/USNO/astrometry/optical-IR-prod/wds/orb6}{ORB6} & $6.89$ & $2000.98$ & $0.773$ & $241$ & $283.8$ & $149.0$ & $49.4$ &-- &-- &--&1&Mason(\citeyear{Mason2005})\\
&\href{https://sb9.astro.ulb.ac.be/mainform.cgi}{SB9}& $6.9185$-Fixed & $1987.187 \pm 0.011$ & $0.77$-Fixed & -- & $109.0 \pm 3.1$ & --& --& $8.83 \pm 0.22$ & $10.90 \pm 0.59$ & $16.44\pm 0.43 $&--&Duquennoy(\citeyear{Duque1991})\\
02128$-$0224&This paper &$0.259516$ &$1983.85112$ &$0.6917$ &$13.82$ &$76.28$ &$240.2$ &$23.9$ &$-5.95$ &$19.01$ &$19.65$& & \\
\textbf{13612} & & $0.259516_{-0.000011}^{+0.000011}$ &$1983.85112_{-0.00026}^{+0.00027}$ &$0.6920_{-0.0027}^{+0.0027}$ &$13.98_{-0.64}^{+0.75}$ &$76.23_{-0.41}^{+0.41}$ &$240.3_{-1.3}^{+1.4}$ &$25.6_{-7.8}^{+6.4}$ &$-5.95_{-0.02}^{+0.02}$ &$19.01_{-0.04}^{+0.04}$ &$19.74_{-0.43}^{+0.40}$ & & \\
& \href{https://www.usno.navy.mil/USNO/astrometry/optical-IR-prod/wds/orb6}{ORB6} & -- & -- & -- &--& -- & -- & -- &-- &-- &--&--&--\\
&\href{https://sb9.astro.ulb.ac.be/mainform.cgi}{SB9}& $0.259515\pm0.000012$ & $1989.55859 \pm 0.00015$ & $0.689\pm0.003$ & -- & $74.3 \pm 0.5$ & --& --& $-5.95 \pm 0.07$ & $19.08 \pm 0.15$ & $19.77\pm 0.15 $&--&Duquennoy(\citeyear{Duque1991})\\
04107$-$0452 & This paper &$20.6327$ &$1931.498$ &$0.8349$ &$164.32$ &$67.88$ &$154.10$ &$67.80$ &$26.49$ &$11.65$ &$12.41$& &\\
\textbf{26441} & & $20.6290_{-0.0076}^{+0.0077}$ &$1931.513_{-0.031}^{+0.031}$ &$0.8384_{-0.0011}^{+0.0011}$ &$164.72_{-0.66}^{+0.66}$ &$67.94_{-0.18}^{+0.18}$ &$154.20_{-0.28}^{+0.28}$ &$67.81_{-0.24}^{+0.23}$ &$26.49_{-0.00}^{+0.00}$ &$11.61_{-0.04}^{+0.04}$ &$12.41_{-0.06}^{+0.06}$ & &\\
& \href{https://www.usno.navy.mil/USNO/astrometry/optical-IR-prod/wds/orb6}{ORB6} & $20.621\pm0.005 $ & $2014.031\pm0.003$ & $0.8400\pm0.0014$ & $167.2\pm2.1$ & $68.6\pm0.3$ & $153.3\pm0.6$ & $69.0\pm0.5$ & &-- &--&2&Tokovinin(\citeyear{Tok2017})\\
& \href{https://sb9.astro.ulb.ac.be/mainform.cgi}{SB9} & $20.608 \pm 0.011$ & $1993.426 \pm 0.011$ & $0.8372\pm0.0015$ & --& $69.0 \pm 0.4$ & --& --& $26.55 \pm 0.03$ & $11.59 \pm 0.06$ & $12.51\pm 0.10 $&5&Griffin(\citeyear{Griffin2015})\\
04184$+$2135& This paper &$11.3630$ &$1955.100$ &$0.1537$ &$135.39$ &$162.61$ &$352.25$ &$123.88$ &$37.79$ &$7.46$ &$8.97$ & &\\
\textbf{27176} & & $11.3642_{-0.0071}^{+0.0071}$ &$1955.100_{-0.030}^{+0.031}$ &$0.1540_{-0.0016}^{+0.0016}$ &$135.37_{-0.16}^{+0.16}$ &$162.67_{-0.80}^{+0.80}$ &$352.22_{-0.24}^{+0.24}$ &$123.88_{-0.25}^{+0.25}$ &$37.79_{-0.01}^{+0.01}$ &$7.50_{-0.32}^{+0.32}$ &$8.96_{-0.11}^{+0.11}$ & &\\
& \href{https://www.usno.navy.mil/USNO/astrometry/optical-IR-prod/wds/orb6}{ORB6} & $11.350\pm0.021 $ & $1977.740\pm0.056$ & $0.1670\pm0.0044$ & $132.90 \pm 0.95$ & $339.\pm1.9$ & $350.70\pm0.61$ & $125.50\pm0.73$ &-- &-- &--&2&Pourbaix(\citeyear{Pour2000})\\
& \href{https://sb9.astro.ulb.ac.be/mainform.cgi}{SB9} & $11.323 \pm 0.016$ & $1989.075 \pm 0.038$ & $0.1711\pm0.0028$ & --& $159.1 \pm 1.3$ & --& --& $37.78 \pm 0.12$ & $7.32 \pm 0.48$ & $9.01\pm 0.16$&--&Torres(\citeyear{Torres1997})\\
07518$-$1354 & This paper & $23.3225$ &$1892.630$ &$0.7537$ &$611.9$ &$253.174$ &$103.054$ &$80.839$ &$-21.37$ &$9.76$ &$9.75$&&\\ \textbf{64096} &
&$23.3207_{-0.0053}^{+0.0054}$ &$1892.639_{-0.027}^{+0.026}$ &$0.7538_{-0.0015}^{+0.0015}$ &$612.0_{-1.6}^{+1.7}$ &$253.185_{-0.079}^{+0.080}$ &$103.044_{-0.061}^{+0.059}$ &$80.835_{-0.044}^{+0.045}$ &$-21.38_{-0.02}^{+0.02}$ &$9.70_{-0.15}^{+0.15}$ &$9.69_{-0.46}^{+0.47}$
&&\\
& \href{https://www.usno.navy.mil/USNO/astrometry/optical-IR-prod/wds/orb6}{ORB6} & $23.330\pm0.010 $ & $1985.923\pm0.011$ & $0.7647\pm0.0021$ & $617.9\pm2.4$ & $253.64\pm0.12$ & $282.65\pm0.09$ & $80.82\pm0.06$ &-- &-- &--&1&Tokovinin(\citeyear{Tok2012b})\\
& \href{https://sb9.astro.ulb.ac.be/mainform.cgi}{SB9} & $22.701 \pm 0.027$ & $1985.914 \pm 0.020$ & $0.741\pm0.007$ & --& $253.1 \pm 0.4$ & --& --& $-21.34\pm 0.16 $ & $9.13 \pm 0.63$ &$9.69 \pm 0.26$ &5&Pourbaix(\citeyear{Pour2000})\\
11560$+$3520& This paper &$13.553$ &$1996.16$ &$0.1010$ &$127.7$ &$316.3$ &$279.5$ &$145.4$ &$-4.99$ &$4.19$ &$4.99$& &\\
\textbf{103613} & & $13.566_{-0.056}^{+0.057}$ &$1996.11_{-0.19}^{+0.18}$ &$0.1008_{-0.0054}^{+0.0054}$ &$127.6_{-2.0}^{+2.0}$ &$315.3_{-4.0}^{+4.0}$ &$279.9_{-1.3}^{+1.3}$ &$146.1_{-3.3}^{+3.7}$ &$-4.99_{-0.01}^{+0.01}$ &$4.19_{-0.12}^{+0.12}$ &$4.98_{-0.02}^{+0.02}$ & &\\
& \href{https://www.usno.navy.mil/USNO/astrometry/optical-IR-prod/wds/orb6}{ORB6} & $13.53 $ & $2010.54$ & $0.165$ & $143$ & $9.90$ & $302.5$ & $135.1$ &-- &-- &--&3&Mason(\citeyear{2013IAUDS3})\\
& \href{https://sb9.astro.ulb.ac.be/mainform.cgi}{SB9} & $13.66 \pm 0.15$ & $2009.46 \pm  0.30$ & $0.103\pm0.011$ & --& $309 \pm 8$ & --& --& $-4.97 \pm 0.04$ & $5.00 \pm 0.06$ & $4.25\pm 0.16 $&5&Griffin(\citeyear{Griff2013})\\
14492$+$1013& This paper&$10.0052$ &$1988.130$ &$0.4994$ &$125.174$ &$162.22$ &$319.48$ &$44.44$ &$-88.29$ &$6.66$ &$6.61$& &\\
\textbf{130669} & & $10.0062_{-0.0042}^{+0.0041}$ &$1988.130_{-0.014}^{+0.014}$ &$0.4999_{-0.0021}^{+0.0022}$ &$125.07_{-0.35}^{+0.35}$ &$162.31_{-0.68}^{+0.67}$ &$319.43_{-0.47}^{+0.47}$ &$44.30_{-0.48}^{+0.48}$ &$-88.29_{-0.01}^{+0.01}$ &$6.65_{-0.06}^{+0.06}$ &$6.61_{-0.10}^{+0.09}$ & &\\
& \href{https://www.usno.navy.mil/USNO/astrometry/optical-IR-prod/wds/orb6}{ORB6} & $10.010 $ & $2018.174$ & $0.518$ & $123$ & $335.9$ & $144.8$ & $43.5$ &-- &-- &--&1&Docobo(\citeyear{2018IAUDS})\\
& \href{https://sb9.astro.ulb.ac.be/mainform.cgi}{SB9} & $9.91 \pm 0.18$ & $2008.127 \pm 0.030$ & $0.488\pm0.009$ & --& $163.0 \pm 2.0$ & --& --& $-88.27 \pm 0.05$ & $6.73 \pm 0.07$ & $6.76\pm 0.09 $&5&Griffin(\citeyear{Griffin2015})\\
15282$-$0921&This paper &$2.4359$ &$1980.4630$ &$0.97525$ &$104.3$ &$254.898$ &$272.47$ &$54.3$ &$7.16$ &$37.45$ &$55.65$& &\\
\textbf{137763} & & $2.4359_{-0.0001}^{+0.0001}$ &$1980.4629_{-0.0002}^{+0.0002}$ &$0.97530_{-0.00011}^{+0.00011}$ &$104.7_{-2.4}^{+2.4}$ &$254.888_{-0.059}^{+0.058}$ &$272.64_{-0.64}^{+0.64}$ &$54.5_{-1.2}^{+1.1}$ &$7.16_{-0.01}^{+0.01}$ &$37.47_{-0.06}^{+0.06}$ &$55.56_{-0.63}^{+0.63}$ & &\\
& \href{https://www.usno.navy.mil/USNO/astrometry/optical-IR-prod/wds/orb6}{ORB6} & $2.43623 $ & $2007.2603$ & $0.976$ & $107.4\pm8.1$ & $255.6$ & $272.8\pm2.1$ & $55.4\pm3.6$ &-- &-- &--&2&Tokovinin(\citeyear{Tok2016})\\
&\href{https://sb9.astro.ulb.ac.be/mainform.cgi}{SB9} & $ 2.436175 \pm 0.000047$ & $1990.205532 \pm 0.000041$ & $0.9733\pm0.0006$ & --& $252.64 \pm 0.73$ & --& --& $7.47 \pm 0.20$ & $36.42 \pm 0.38$ & $52.90\pm 1.73 $&--&Halbwachs(\citeyear{Hal2018})\\
16584$+$3943&This paper &$16.023$ &$1985.170$ &$0.7732$ &$75.40$ &$336.04$ &$245.46$ &$109.38$ &$33.54$ &$10.46$ &$11.90$& &\\
\textbf{153527} & & $16.022_{-0.039}^{+0.039}$ &$1985.171_{-0.040}^{+0.040}$ &$0.7729_{-0.0019}^{+0.0019}$ &$75.38_{-0.42}^{+0.41}$ &$336.06_{-0.43}^{+0.43}$ &$245.36_{-0.46}^{+0.47}$ &$109.55_{-0.82}^{+0.83}$ &$33.54_{-0.08}^{+0.08}$ &$10.45_{-0.09}^{+0.09}$ &$11.89_{-0.13}^{+0.13}$ & &\\
& \href{https://www.usno.navy.mil/USNO/astrometry/optical-IR-prod/wds/orb6}{ORB6} & $16.09 $ & $2001.07$ & $0.753$ & $75$ & $154.3$ & $63.2$ & $113.0$ &-- &-- &--&2&Docobo(\citeyear{2013IAUDS})\\
&\href{https://sb9.astro.ulb.ac.be/mainform.cgi}{SB9} & $16.153$-Fixed & $ 2001.1970\pm 0.0080$ & $0.775\pm0.003$ & --& $337.0 \pm 0.9$ & --& --& $33.59 \pm 0.16$ & $10.40 \pm 0.18$ & $11.96\pm 0.23 $&5&Griffin(\citeyear{Griff2003})\\
18384$-$0312&This paper &$12.1422$ &$1910.000$ &$0.2454$ &$145.35$ &$257.88$ &$352.99$ &$123.90$ &$-15.97$ &$7.21$ &$7.55$& &\\
\textbf{172088} & & $12.1426_{-0.0044}^{+0.0044}$ &$1909.996_{-0.039}^{+0.039}$ &$0.2456_{-0.0019}^{+0.0019}$ &$145.37_{-0.39}^{+0.38}$ &$257.87_{-0.30}^{+0.30}$ &$353.03_{-0.22}^{+0.22}$ &$123.87_{-0.22}^{+0.22}$ &$-15.97_{-0.00}^{+0.00}$ &$7.20_{-0.03}^{+0.03}$ &$7.55_{-0.03}^{+0.03}$ & &\\
& \href{https://www.usno.navy.mil/USNO/astrometry/optical-IR-prod/wds/orb6}{ORB6} & $12.135 \pm 0.006 $ & $2007.18 \pm 0.04$ & $0.2567 \pm 0.0029$ & $146.1\pm0.8$ & $79.7\pm1.5$ & $173.4\pm 0.9$ & $123.9\pm5.0 $ &-- &-- &--&1&Hartkopf(\citeyear{Griffin2013})\\
&\href{https://sb9.astro.ulb.ac.be/mainform.cgi}{SB9} & $12.1342$-Fixed & $ 2007.185\pm 0.049$ & $0.230\pm0.006$ & --& $258.9 \pm 1.5$ & --& --& $-15.95 \pm 0.03$ & $7.19 \pm 0.06$ & $7.51\pm 0.06 $&5&Griffin(\citeyear{Griffin2013})\\
20102$+$4357&This paper &$85.05$ &$1886.15$ &$0.4945$ &$448.6$ &$344.00$ &$143.75$ &$117.15$ &$-42.99$ &$3.51$ &$5.40$& &\\
\textbf{191854} & & $85.02_{-0.27}^{+0.27}$ &$1886.22_{-0.28}^{+0.27}$ &$0.4889_{-0.0035}^{+0.0035}$ &$448.2_{-1.3}^{+1.34}$ &$344.07_{-0.82}^{+0.82}$ &$143.75_{-0.26}^{+0.25}$ &$117.22_{-0.28}^{+0.28}$ &$-43.06_{-0.14}^{+0.13}$ &$3.65_{-0.36}^{+0.37}$ &$5.07_{-0.49}^{+0.47}$ & &\\
& \href{https://www.usno.navy.mil/USNO/astrometry/optical-IR-prod/wds/orb6}{ORB6} & $85.61 $ & $1970.27$ & $0.488 $ & $449 $ & $339.4$ & $142.2$ & $116.4 $ &-- &-- &--&2&Heintz(\citeyear{Heintz1997})\\
&\href{https://sb9.astro.ulb.ac.be/mainform.cgi}{SB9} & $85.22\pm0.12$ & $ 1970.44\pm0.18$ & $0.4922\pm0.0033$ & --& $339 \pm 1 $ & --& --& $-43.2 \pm 0.2$ & $3.97 \pm 0.44$ & $5.12\pm 0.47$&--&Pourbaix(\citeyear{Pour2000})\\
20205$+$4351&This paper &$7.9232$ &$1993.1840$ &$0.9060$ &$8.44$ &$42.76$ &$347.7$ &$117.5$ &$1.43$ &$28.0$ &$68.1$& &\\
\textbf{193793} & & $7.9241_{-0.0035}^{+0.0037}$ &$1993.1825_{-0.0039}^{+0.0036}$ &$0.9012_{-0.0034}^{+0.0034}$ &$8.30_{-0.23}^{+0.22}$ &$42.31_{-0.68}^{+0.70}$ &$349.1_{-2.5}^{+2.5}$ &$119.4_{-3.4}^{+3.7}$ &$1.25_{-0.05}^{+0.05}$ &$28.3_{-1.8}^{+1.8}$ &$68.5_{-1.1}^{+1.2}$ & &\\
& \href{https://www.usno.navy.mil/USNO/astrometry/optical-IR-prod/wds/orb6}{ORB6} & $7.9298 \pm 0.0025 $ & $1985.2443\pm0.0082$ & $0.901\pm0.005 $ & $8.99\pm0.19 $ & $48.2\pm1.9$ & $354.2\pm0.7 $ & $118.9\pm0.9 $ &-- &-- &--&3&Monnier(\citeyear{Monnier2011})\\
&\href{https://sb9.astro.ulb.ac.be/mainform.cgi}{SB9} & $7.9370 \pm 0.0036 $ & $ 1985.2225\pm 0.0044 $ & $0.881\pm0.005$ & --& $46.7\pm1.6 $ & --& --& $3.1 \pm 1.0$ & $30.5 \pm 1.9$ & $82.0\pm 2.3$&--&Marchenko(\citeyear{March2003})\\
20527$+$4607&This paper &$31.141$ &$1918.97$ &$0.7505$ &$236.7$ &$321.90$ &$47.72$ &$127.52$ &$-20.30$ &$6.95$ &$7.54$& &\\
\textbf{--} & & $31.144_{-0.051}^{+0.052}$ &$1918.97_{-0.11}^{+0.10}$ &$0.7510_{-0.0034}^{+0.0034}$ &$236.4_{-2.0}^{+2.0}$ &$322.01_{-0.66}^{+0.67}$ &$47.74_{-0.54}^{+0.53}$ &$127.57_{-0.86}^{+0.89}$ &$-20.31_{-0.04}^{+0.04}$ &$6.96_{-0.10}^{+0.10}$ &$7.54_{-0.10}^{+0.10}$ &  &\\
& \href{https://www.usno.navy.mil/USNO/astrometry/optical-IR-prod/wds/orb6}{ORB6} & $30.45 $ & $2013.49 $ & $0.723 $ & $212$ & $351.6$ & $65.9$ & $128.7 $ &-- &-- &--&3&Docobo(\citeyear{2013IAUDS2})\\
&\href{https://sb9.astro.ulb.ac.be/mainform.cgi}{SB9} & $31.4853 $ & $ 1981.2329$ & $0.74$ & --& $320. $ & --& --& $-20.3$ & $6.9$ & $7.5$&2&Griffin(\citeyear{Griff1984})\\
23485$+$2539& This paper&$3.2163$ &$2001.3988$ &$0.6061$ &$40.3$ &$77.75$ &$121.8$ &$86.1$ &$-9.55$ &$16.43$ &$16.35$& &\\
\textbf{223323} & & $3.2157_{-0.0035}^{+0.0035}$ &$2001.3995_{-0.0042}^{+0.0041}$ &$0.6060_{-0.0038}^{+0.0038}$ &$40.3_{-1.2}^{+1.2}$ &$77.79_{-0.60}^{+0.60}$ &$121.2_{-1.7}^{+1.7}$ &$86.0_{-1.3}^{+1.3}$ &$-9.55_{-0.01}^{+0.01}$ &$16.43_{-0.12}^{+0.12}$ &$16.35_{-0.12}^{+0.12}$ & &\\
& \href{https://www.usno.navy.mil/USNO/astrometry/optical-IR-prod/wds/orb6}{ORB6} & $3.2191 \pm 0.0038  $ & $2011.0545 \pm 0.0085$ & $0.602 \pm 0.008$ & $40.8\pm1.0 $ & $78.4\pm0.6$ & $121.1\pm 1.1$ & $86.0\pm0.9 $ &-- &-- &--&3&Horch(\citeyear{Horch2019})\\
&\href{https://sb9.astro.ulb.ac.be/mainform.cgi}{SB9} & $3.2172 \pm 0.0036$ & $ 2004.6134 \pm 0.0025$ & $0.604\pm0.003$ & --& $77.8 \pm 0.6$ & --& --& $-9.56 \pm 0.06$ & $16.38 \pm 0.11$ & $16.40\pm 0.16$&--&Griffin(\citeyear{Griff2007})\\
\enddata
\end{deluxetable}

\begin{figure} 
      \begin{minipage}[!h]{0.5\linewidth}
        \centering
        \includegraphics[height=.25\textheight]{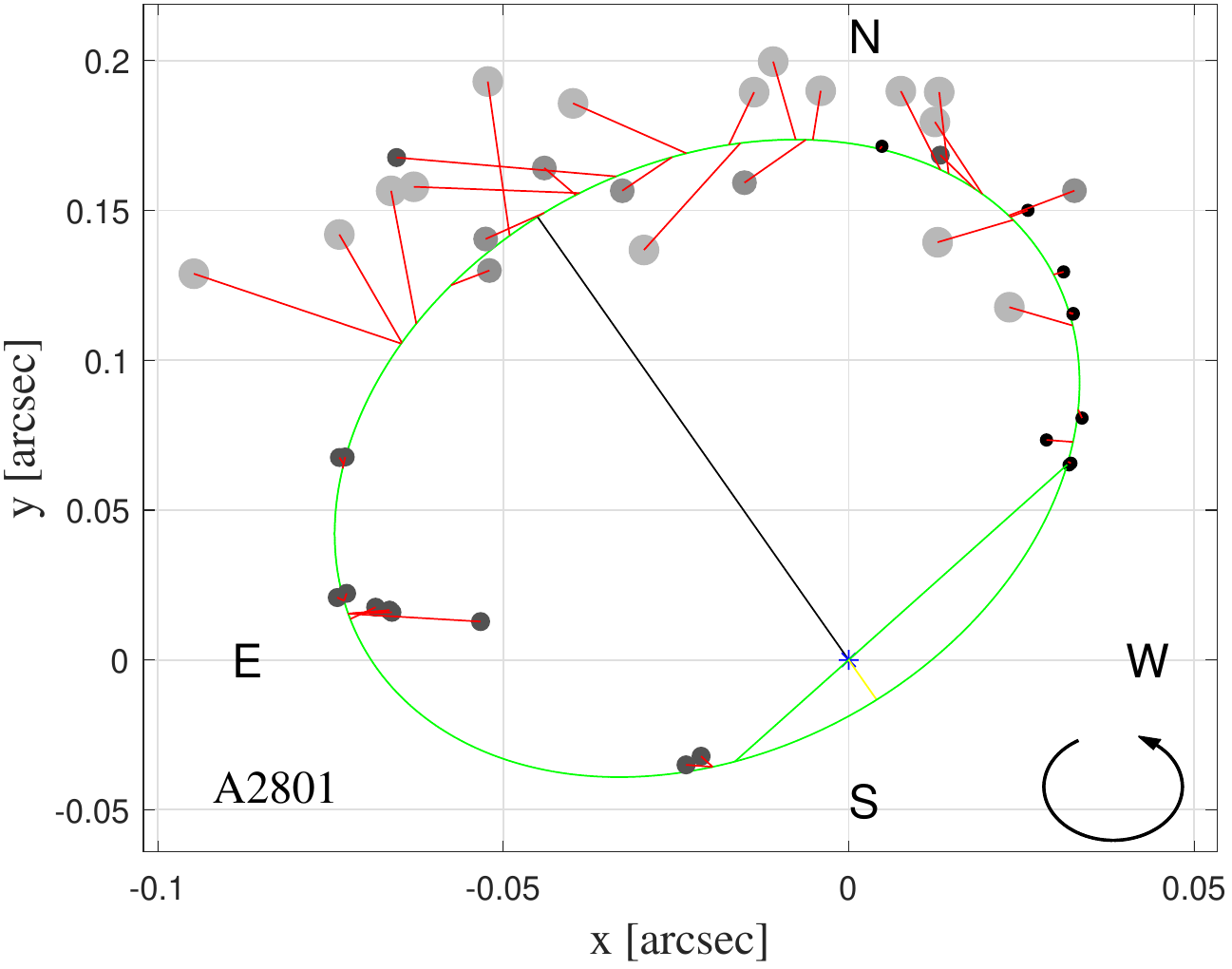}
        \includegraphics[height=.25\textheight]{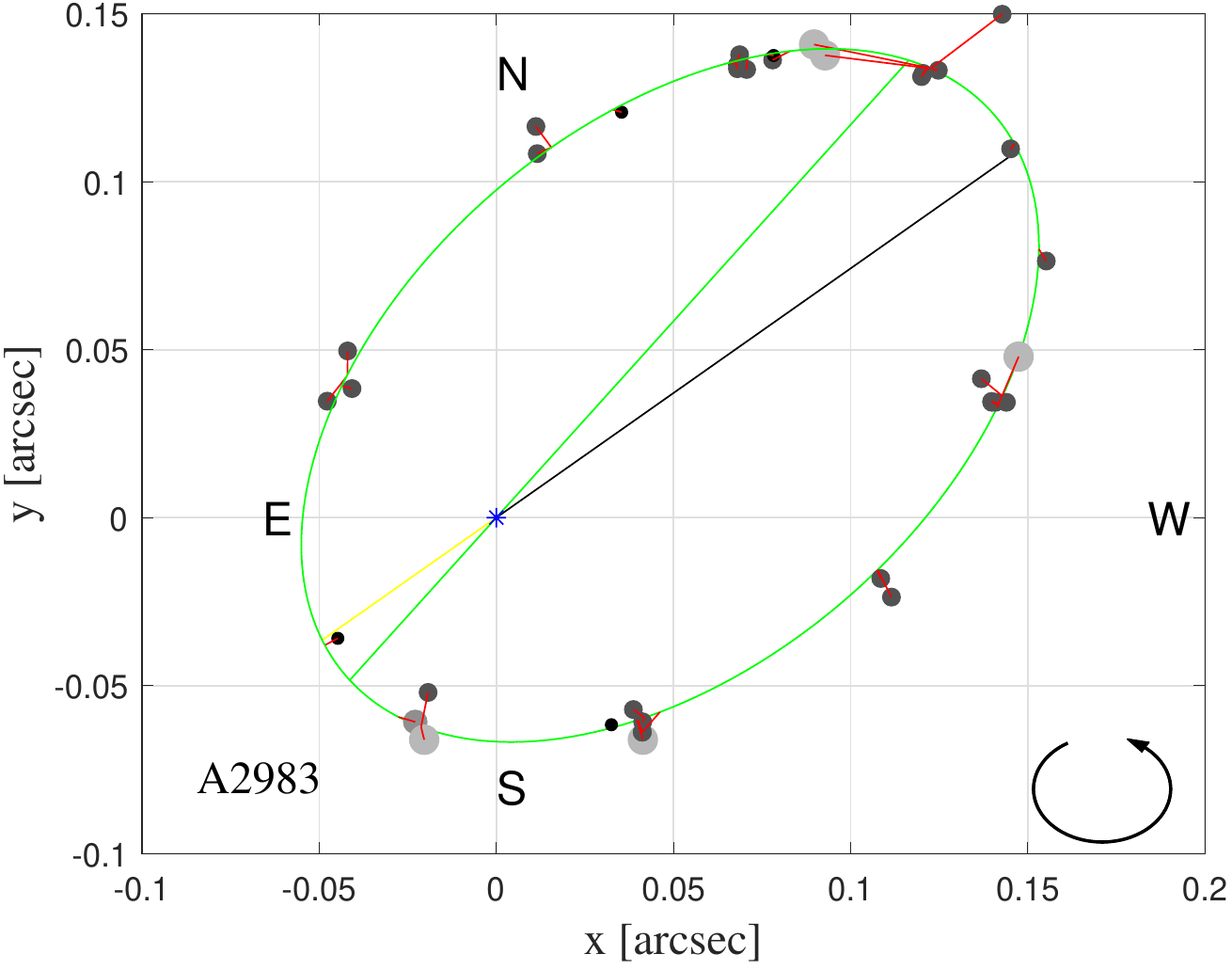}
        \includegraphics[height=.25\textheight]{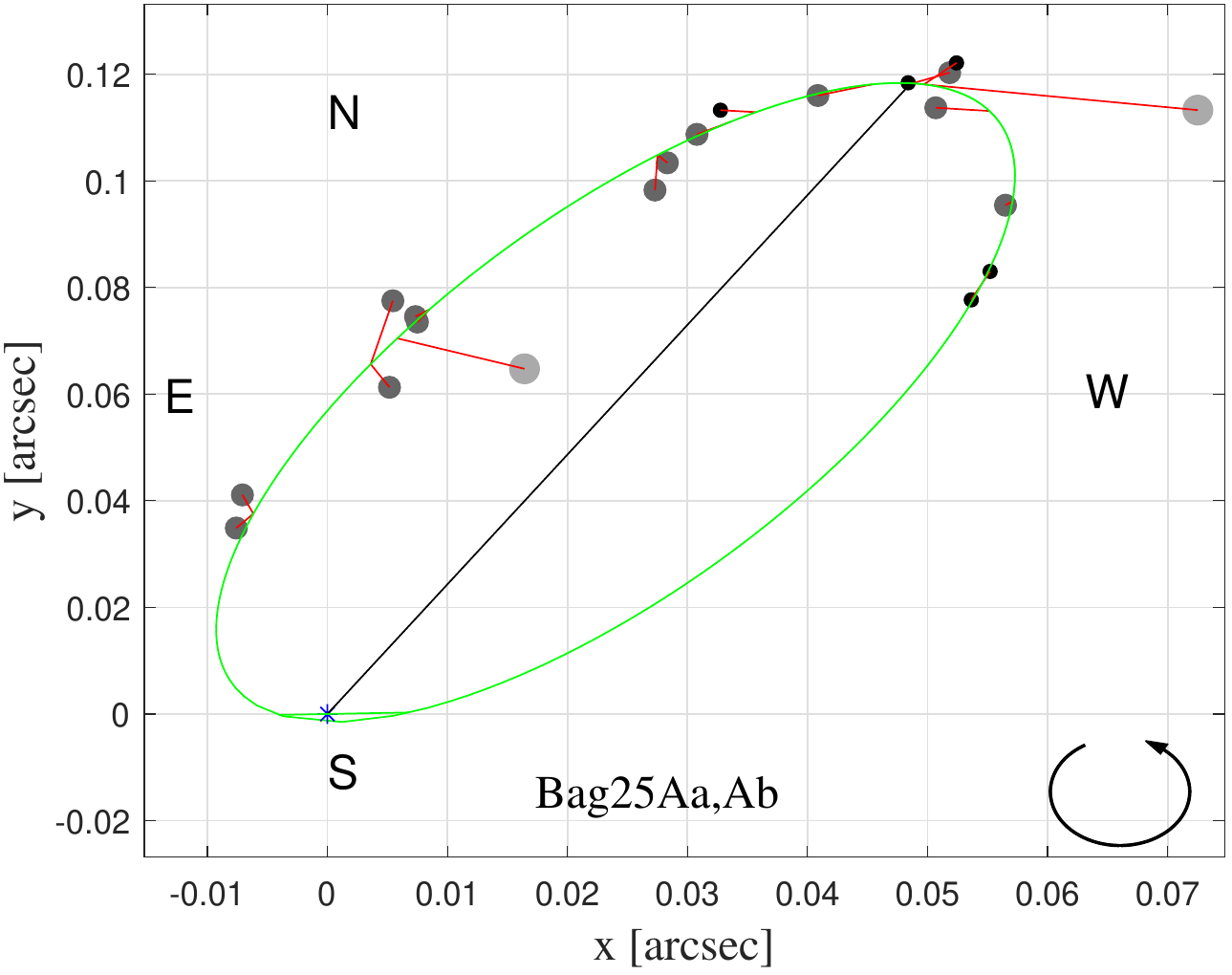}
      \end{minipage}
      \begin{minipage}[!h]{0.5\linewidth}
        \centering
        \includegraphics[height=.25\textheight]{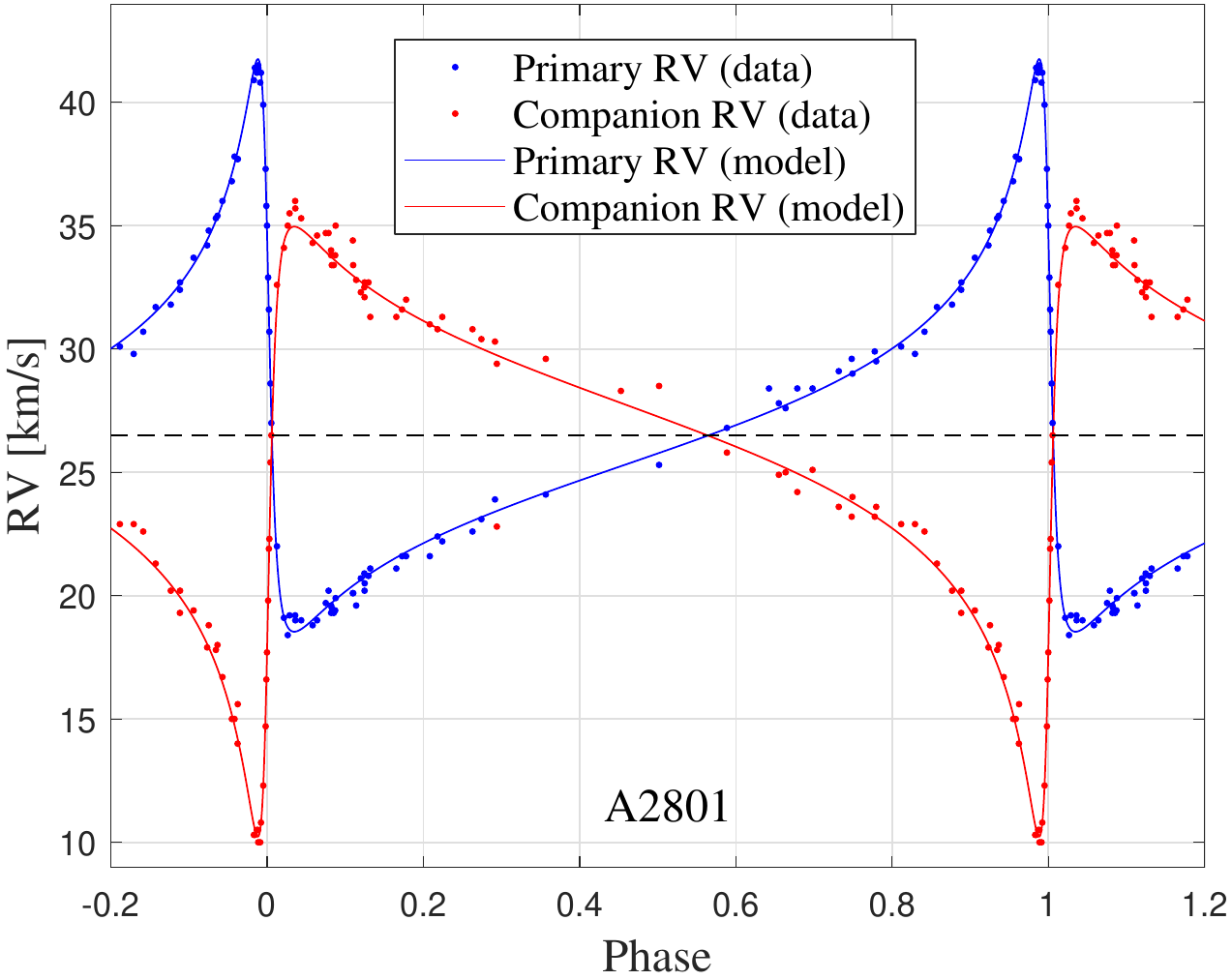}
        \includegraphics[height=.25\textheight]{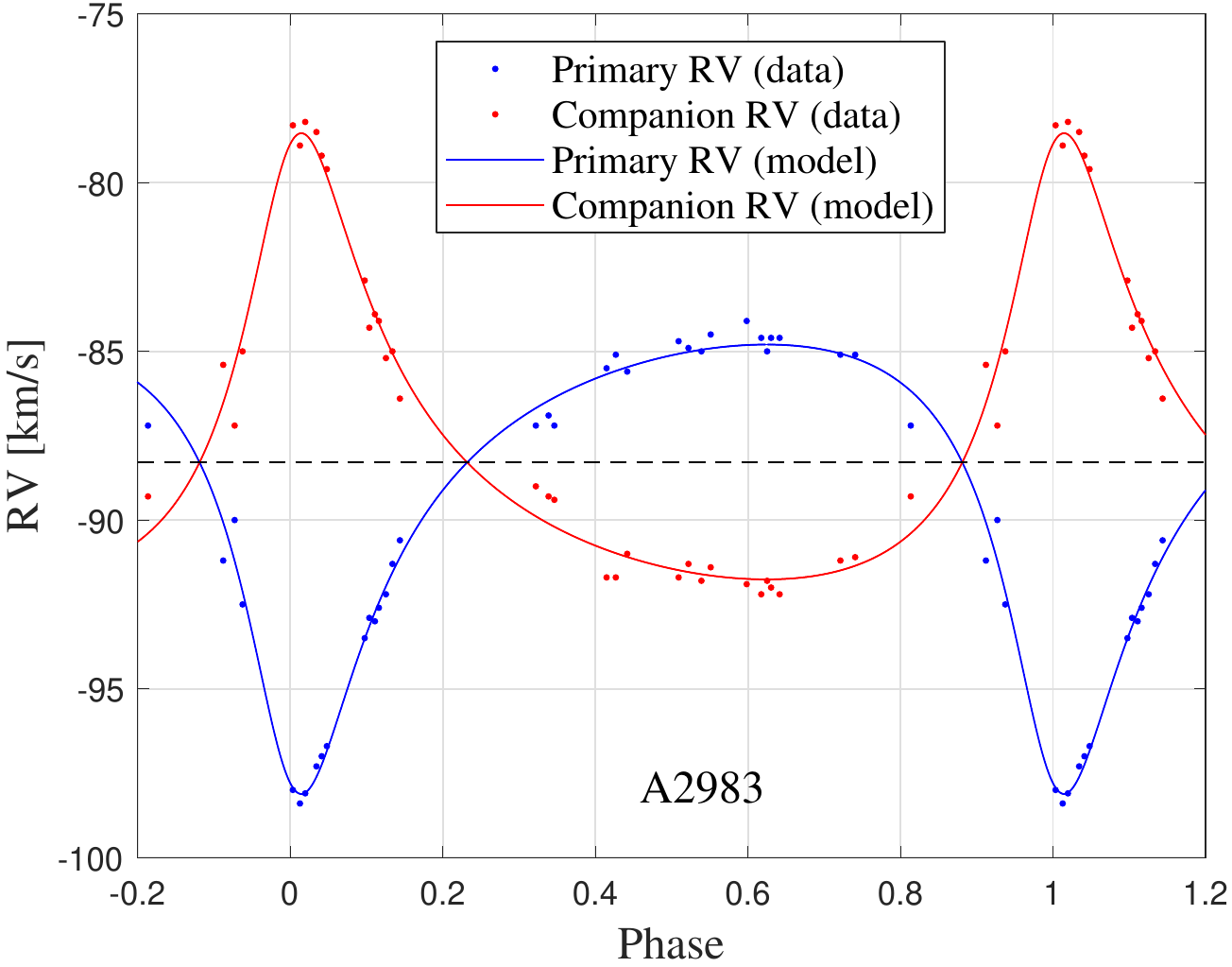}
        \includegraphics[height=.25\textheight]{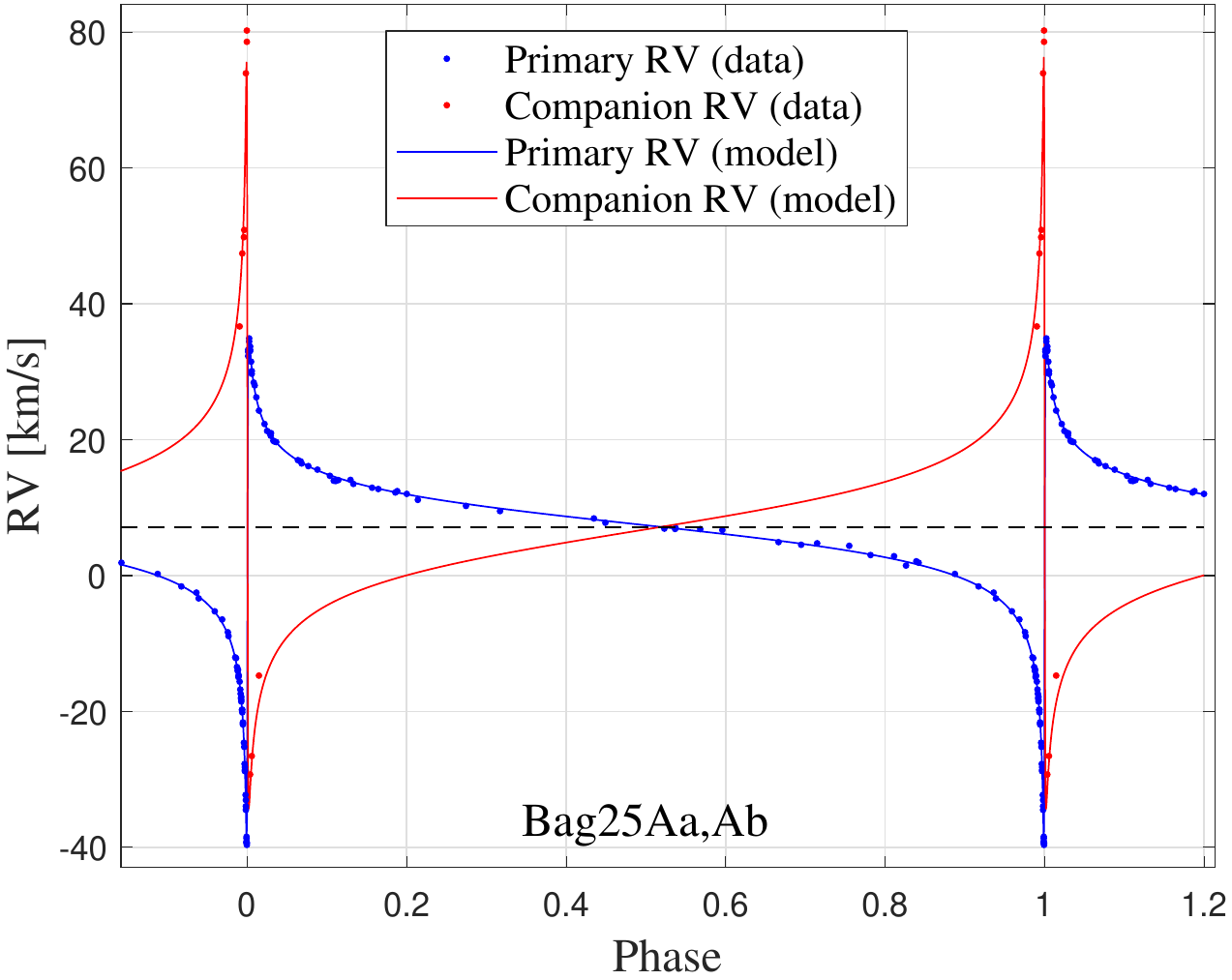}
      \end{minipage} 
\caption{Maximum likelihood orbits from simultaneous fits to the astrometric and RV curves for three representative cases. From top to bottom: A2801 (WDS 04107-0452), A2983 (WDS 14492+1013), and Bag25Aa,Ab (WDS 15282-0921). The left panel shows the data points and the astrometric orbit. The size and color of the dots indicate the weight (uncertainty) of each observation: large clear dots indicate larger errors and the opposite is true for small dark dots. Smaller dots are from more recent interferometric measurements, including - but not limited to - our own. The green line indicates the line of nodes, while the black line indicates the direction to apoastron. The right panel shows the RV curves of both components. The horizontal dashed line indicates the inferred (fitted) systemic velocity reported in Table~\ref{tab:orbel}.\label{fig:examples}}
\end{figure}

\begin{figure}[!ht]
\centering
\boxedfig{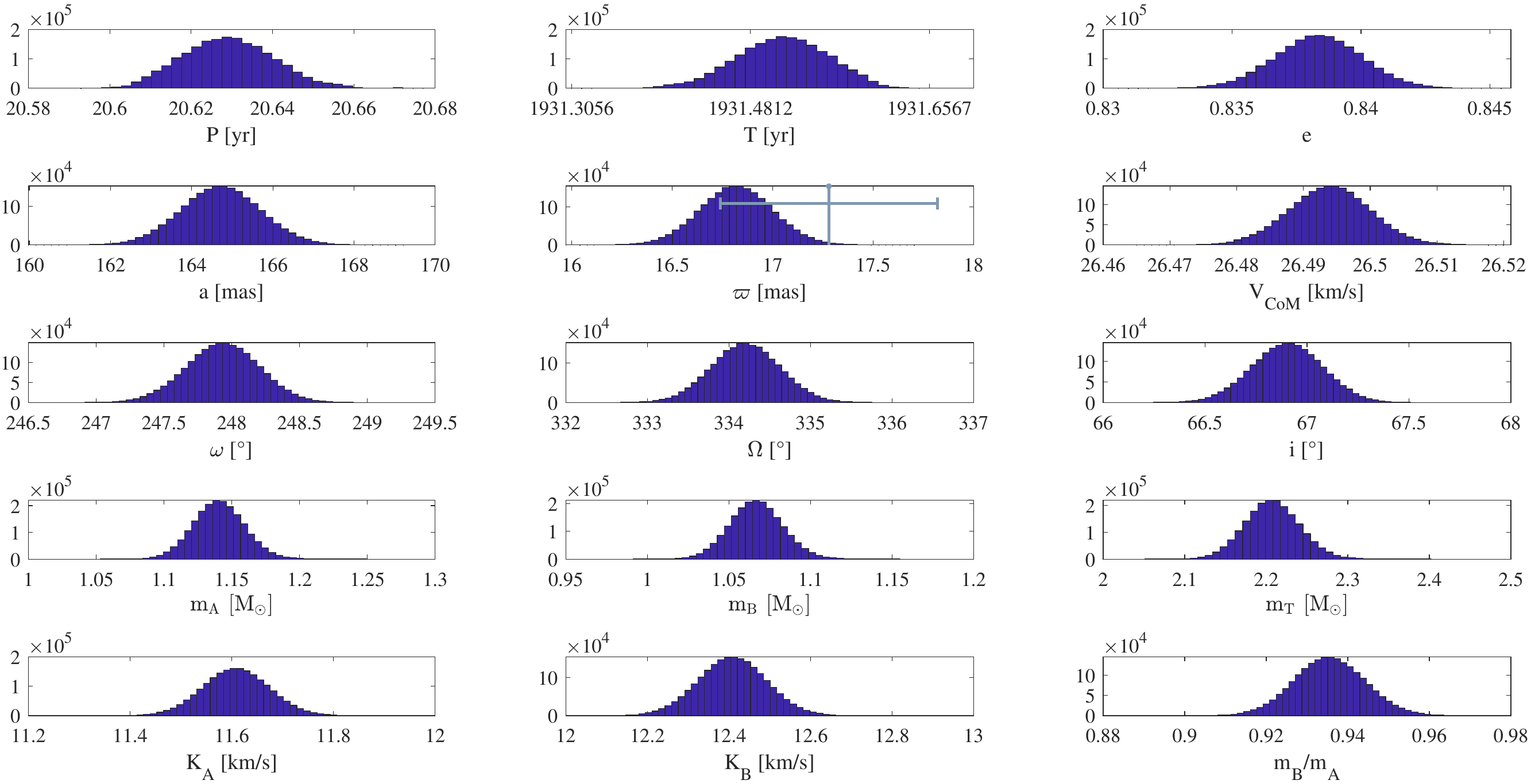}{0.72\textwidth}{\textbf{A2801}}\\
\boxedfig{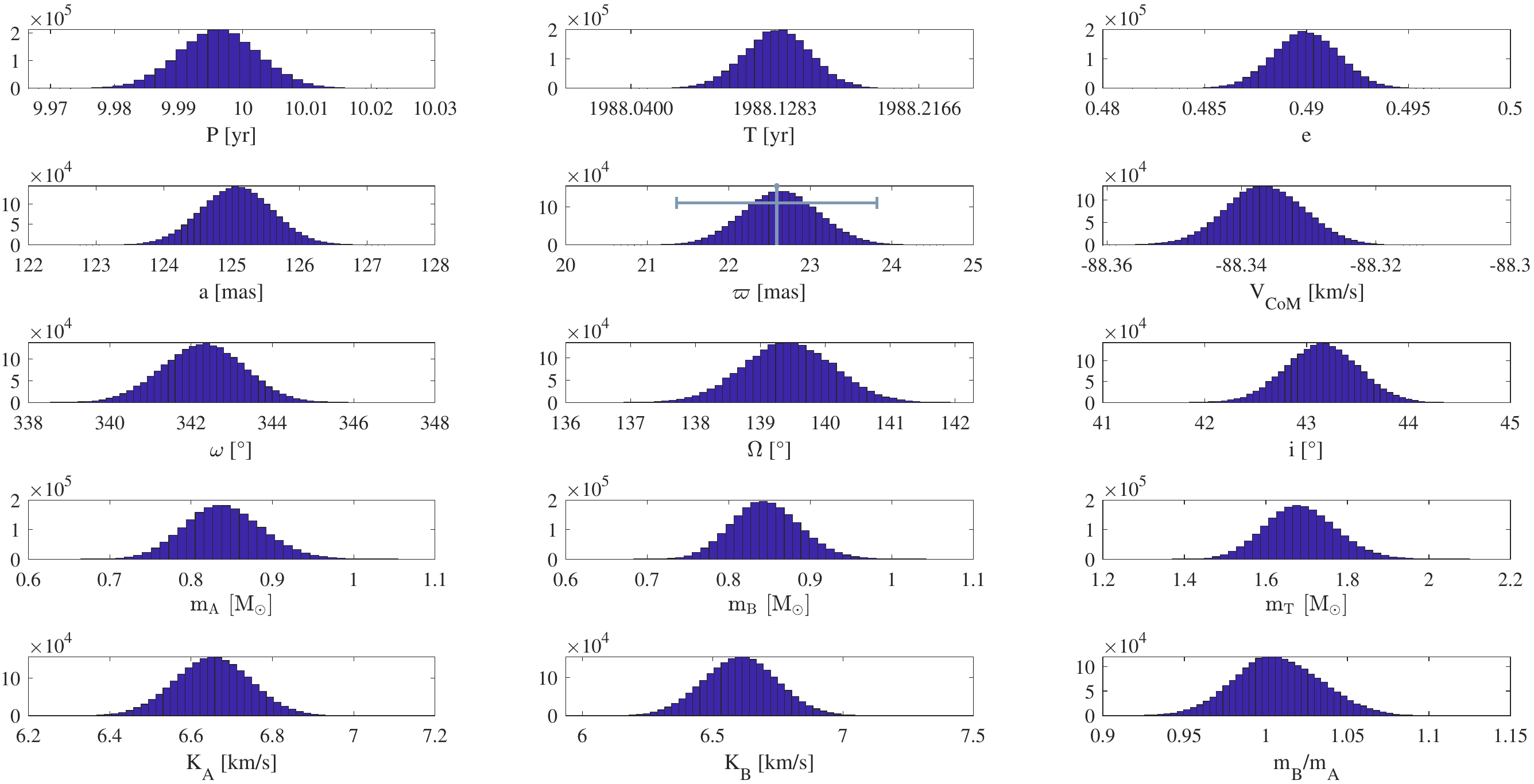}{0.72\textwidth}{\textbf{A2983}}\\
\boxedfig{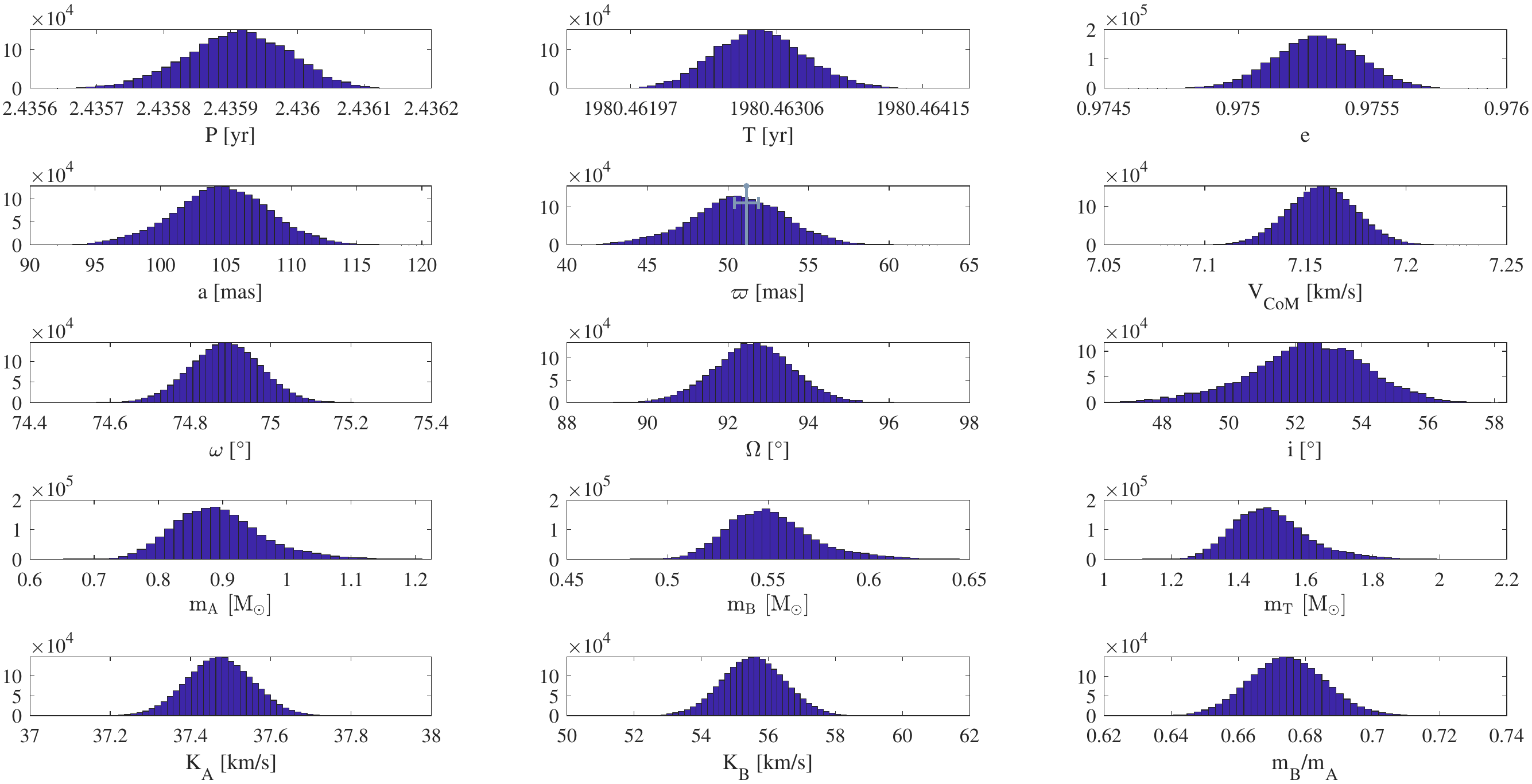}{0.72\textwidth}{\textbf{BAG25Aa,Ab}}\\
\caption{Posterior distributions of the classical seven orbital elements; plus the (fitted) orbital parallax, the Gaia DR2 trigonometric parallax and its $\pm 1 \sigma$ error, the systemic velocity, the velocity amplitudes for both components, the mass sum, the mass ratio and the individual component masses, for the same objects shown in Figure~\ref{fig:examples}. We note that for objects with a mass-ratio close to one (in this case A2983 - middle panel; but see Table~\ref{tab:plx} for other objects) the $m_{\mbox{B}}/m_{\mbox{A}}$ histograms are well-behaved and smooth across that boundary, as explained in the text.}\label{fig:hist}
\end{figure}

In Figure~\ref{fig:examples} we show a subset of representative examples of orbital solutions from our simultaneous fit to the astrometric orbit (left panel) and the RV curve (right panel); and in Figure~\ref{fig:hist} we present the PDFs for the same three systems. Inspection of Table~\ref{tab:orbel}, and Figures~\ref{fig:examples} and \ref{fig:hist} shows that well determined orbits have a ML value that approximately coincides with the 2nd quartile of the PDF, that their inter-quartile range is relatively well constrained, and that the PDFs have a Gaussian-like distribution. On the contrary, poor orbits show PDFs with long tails -and therefore large inter-quartile ranges- on which the ML value usually differs significantly from the 2nd quartile, and the PDFs are tangled. The upper panel of Figure~\ref{fig:hip10305pdf}) shows an  extreme case in this respect: TOK39Aa,Ab. For completeness, in \url{http://www.das.uchile.cl/~rmendez/B_Research/JAA-RAM-SB2/} we make available the orbital plots and the relevant PDFs for all the systems in our sample. Even though the astrometric orbits do not always have an excellent phase coverage, the combined solution produces very precise orbital parameters. This is most evident in Figure~\ref{fig:hist}, which exhibit tight and well-constrained distributions. Judging from our quartile-based uncertainty estimation, we can see that the mass ratio for this sample of objects is determined in the best cases with less than 1\% error, while the uncertainty on the mass sum is around 1\%. The formal error on the best individual component masses that we could determine is $0.01 M_\odot$ (see Table~\ref{tab:plx}).

\floattable
\begin{deluxetable}{cccccccc}
\tablecaption{Parallaxes and individual component masses. \label{tab:plx}}
\tabletypesize{\footnotesize}
\tablecolumns{3}
\tablewidth{0pt}
\tablehead{\colhead{WDS name} &
\colhead{Hipparcos} &
\colhead{GAIA eDR3} &
\colhead{Orbital} &
\colhead{$m_{\mbox{B}}$/$m_{\mbox{A}}$} &
\colhead{$m_{\mbox{T}}$} & 
\colhead{$m_{\mbox{A}}$} & 
\colhead{$m_{\mbox{B}}$} \\ \colhead{\textbf{HD number}}
& \colhead{(mas)} & \colhead{(mas)} & \colhead{(mas)} &
$~$ & \colhead{$M_{\odot}$} & \colhead{$M_{\odot}$} & \colhead{$M_{\odot}$}}
\startdata
00352$-$0336 & $47.05 \pm 0.67$ & -- & $42.41$ &$0.746$ &$3.54$ &$2.03$ &$1.51$\\
\textbf{3196} & & & $42.38_{-0.85}^{+0.88}$ &$0.749_{-0.034}^{+0.034}$ &$3.54_{-0.21}^{+0.21}$ &$2.02_{-0.09}^{+0.09}$ &$1.52_{-0.12}^{+0.13}$\\
02128$-$0224\tablenotemark{1}&$25.19\pm1.41$ &$26.54\pm0.13$ & $23.1$ &$0.968$ &$3.2$ &$1.62$ &$1.57$\\
\textbf{13612} & & & $24.8_{-8.1}^{+7.2}$ &$0.963_{-0.020}^{+0.022}$ &$2.7_{-1.2}^{+4.9}$
&$1.35_{-0.62}^{+2.49}$ &$1.31_{-0.60}^{+2.42}$\\
02128$-$0224+prior\tablenotemark{1}& $25.19\pm1.41$ &$26.54\pm0.13$ & $26.03$ &$0.959$ &$2.355$ &$1.20$ &$1.15$\\
\textbf{13612} & & & $26.29_{-0.21}^{+0.21}$ &$0.960_{-0.020}^{+0.022}$ &$2.301_{-0.084}^{+0.087}$ &$1.17_{-0.05}^{+0.05}$ &$1.13_{-0.04}^{+0.04}$\\
04107$-$0452 & $16.09 \pm 0.65$ & $17.28 \pm 0.54$ & $16.58$ &$0.9386$ &$2.286$ &$1.18$ &$1.11$\\
\textbf{26441} & & & $16.82_{-0.12}^{+0.12}$ &$0.9355_{-0.0058}^{+0.0059}$ &$2.207_{-0.022}^{+0.023}$ &$1.14_{-0.01}^{+0.01}$ &$1.07_{-0.01}^{+0.01}$\\
04184$+$2135 & $18.5 \pm 0.5$ & $19.51 \pm 0.24$ & $18.15$ &$0.832$ &$3.21$ &$1.76$ &$1.46$\\
\textbf{27176} & & & $18.11_{-0.37}^{+0.39}$ &$0.837_{-0.038}^{+0.037}$ &$3.23_{-0.20}^{+0.20}$ &$1.76_{-0.08}^{+0.08}$ &$1.47_{-0.12}^{+0.12}$\\
07518$-$1354 & $60.59 \pm 0.59$ & -- & $60.2$ &$1.001$ &$1.93$ &$0.97$ &$0.97$\\
\textbf{64096} & & & $60.6_{-1.5}^{+1.6}$ &$1.002_{-0.050}^{+0.053}$ &$1.90_{-0.14}^{+0.15}$ &$0.95_{-0.09}^{+0.10}$ &$0.95_{-0.05}^{+0.06}$\\
11560$+$3520 & $13.86 \pm 0.58$ & $14.73 \pm 0.37$ & $17.4$ &$0.839$ &$2.14$ &$1.16$ &$0.98$\\
\textbf{103613} & & & $17.1_{-1.9}^{+1.6}$ &$0.841_{-0.025}^{+0.026}$ &$2.25_{-0.48}^{+0.83}$ &$1.22_{-0.26}^{+0.45}$ &$1.03_{-0.22}^{+0.38}$\\
14492$+$1013 & $22.6 \pm 1.2$ & -- & $22.69$ &$1.007$ &$1.679$ &$0.84$ &$0.84$\\
\textbf{130669} & & & $22.64_{-0.31}^{+0.31}$ &$1.007_{-0.017}^{+0.019}$&$1.684_{-0.057}^{+0.060}$ &$0.84_{-0.03}^{+0.03}$ &$0.84_{-0.03}^{+0.03}$\\
15282$-$0921 & $48.6 \pm 1.3$ & -- & $50.3$ &$0.6731$ &$1.5024$ &$0.90$ &$0.60$\\
\textbf{137763} & & & $50.7_{-1.9}^{+2.0}$ &$0.6745_{-0.0076}^{+0.0077}$ &$1.486_{-0.068}^{+0.073}$ &$0.89_{-0.04}^{+0.05}$ &$0.60_{-0.03}^{+0.03}$\\
16584$+$3943 & $8.8 \pm 0.68$ & $8.97 \pm 0.05$ & $9.325$ &$0.879$ &$2.059$ &$1.10$ &$0.96$\\
\textbf{153527} & & & $9.313_{-0.099}^{+0.096}$ &$0.879_{-0.016}^{+0.016}$ &$2.065_{-0.036}^{+0.038}$ &$1.10_{-0.02}^{+0.02}$ &$0.97_{-0.02}^{+0.02}$\\
18384$-$0312 & $20.85 \pm 0.91$ & -- & $20.69$ &$0.9546$ &$2.351$ &$1.20$ &$1.15$\\
\textbf{172088} & & & $20.70_{-0.12}^{+0.12}$ &$0.9539_{-0.0056}^{+0.0057}$ &$2.349_{-0.023}^{+0.024}$ &$1.20_{-0.01}^{+0.01}$ &$1.15_{-0.01}^{+0.01}$\\
20102$+$4357 & $19.48 \pm 0.54$ & $19.30 \pm 0.13$ & $18.04$ &$0.65$ &$2.12$ &$1.29$ &$0.84$\\
\textbf{191854} & & & $18.39_{-0.89}^{+0.99}$ &$0.72_{-0.11}^{+0.13}$ &$2.00_{-0.29}^{+0.32}$ &$1.16_{-0.20}^{+0.22}$ &$0.84_{-0.13}^{+0.14}$\\
20205$+$4351 & $0.25 \pm 0.42$ & $0.538 \pm 0.024$ & $0.691$ &$0.411$ &$29.0$ &$20.5$ &$8.4$\\
\textbf{193793} & & & $0.648_{-0.044}^{+0.040}$ &$0.413_{-0.027}^{+0.027}$ &$33.6_{-3.5}^{+4.7}$ &$23.7_{-2.4}^{+3.2}$ &$9.9_{-1.3}^{+1.6}$\\
20527$+$4607 & $19.0 \pm 1.0$ & $18.07 \pm 0.56$ & $18.75$ &$0.922$ &$2.073$ &$1.08$ &1.00\\
-- & & & $18.71_{-0.41}^{+0.40}$ &$0.923_{-0.020}^{+0.020}$ &$2.079_{-0.083}^{+0.090}$ &$1.08_{-0.04}^{+0.05}$ &$1.00_{-0.04}^{+0.04}$\\
23485$+$2539 & $14.51 \pm 0.47$ & $14.42 \pm 0.03$ & $14.28$ &$1.005$ &$2.172$ &$1.08$ &$1.09$\\
\textbf{223323} & & & $14.25_{-0.45}^{+0.45}$ &$1.005_{-0.014}^{+0.014}$ &$2.176_{-0.026}^{+0.026}$ &$1.09_{-0.01}^{+0.01}$ &$1.09_{-0.01}^{+0.01}$\\
\enddata
\tablenotetext{1}{The first solution is computing an orbital parallax. The second solution is imposing the GAIA eDR3 parallax as a (Gaussian) prior to the solution. See also Figure~\ref{fig:hip10305pdf} and Section~\ref{sec:objnotes}}
\end{deluxetable}

In Table~\ref{tab:plx} we present a comparison of the parallax values from Hipparcos and/or Gaia-eDR3 with our orbital parallaxes. In the last four columns we give the  mass ratio, the total mass, and the individual masses obtained from our simultaneous fits to the astrometric and RV data, which was done adopting the orbital elements given in Table~\ref{tab:orbel}). The first lines give the ML values and the second line the quartiles. All mass values were computed allowing the parallax of the system to be a free parameter of the MCMC code, i.e., using the orbital parallax\footnote{Except for target TOK39Aa,Ab=WDS 02128-0224, for which we used a parallax prior; see Section~\ref{sec:objnotes} and Figure~\ref{fig:hip10305pdf}. In fact, in Table~\ref{tab:plx} we report both solutions for this target: with and without a parallax prior.}, whose ML value and quartiles are given in the fourth column of this table (upper and lower row respectively). Therefore, our mass estimates {\it do} include the extra variance from this parameter. 

\begin{figure}[ht]
\centering
\epsscale{1.2}
\plotone{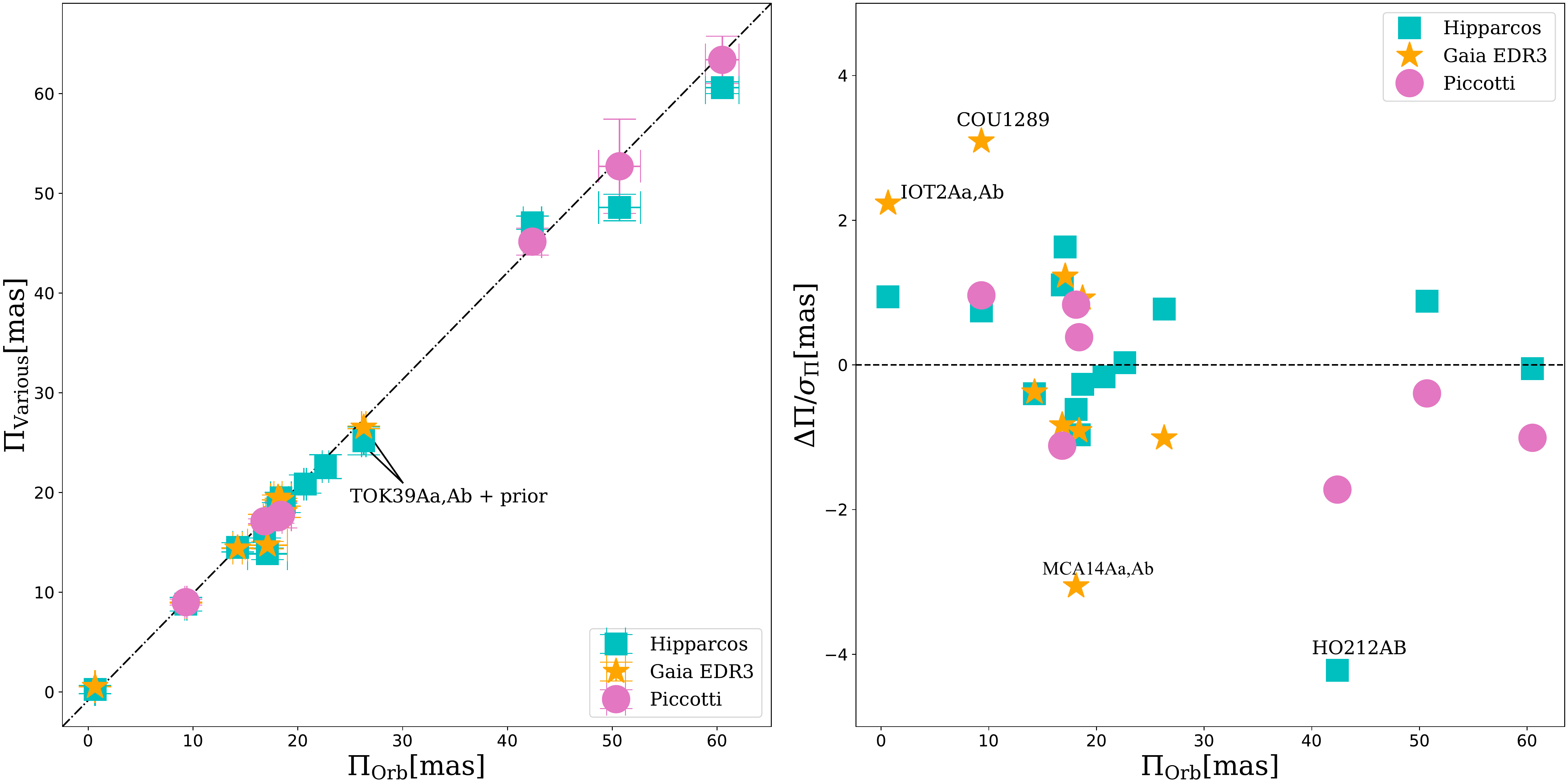}
\caption{Comparison of trigonometric and orbital parallaxes for our sample of SB2 binaries. The left panel shows a comparison of our orbital parallax $\Pi_{\mbox{Orb}}$, with the Hipparcos parallax (from the re-reduction by \citet{vanLee2010})), the Gaia eDR3 parallax and the orbital parallaxes from \citet{Picc2020}. The correlation is good and tight. The dot-dashed line is a one-to-one relationship shown for reference. In the right panel we plot the deviation of $\left( \Pi_{\mbox{Orb}} - \Pi_{\mbox{Various}} \right) / \sigma_{\Pi}$, where $\sigma_{\Pi}$ includes our uncertainty and those quoted for Hipparcos, Gaia and Piccotti in Table~\ref{tab:plx}. The four most discrepant cases, identified in the plot, are further discussed in Section~\ref{sec:objnotes}.}
\label{fig:par}
\end{figure}

In Figure~\ref{fig:par} we show a comparison of our orbital parallaxes with those from Hipparcos, Gaia, and the recent study of SB2 binaries by \citet{Picc2020}. The mean values of the differences $\left< \Pi_{\mbox{orb}} - \Pi_{\mbox{various}} \right>$ are $0.0 \pm 1.8$~mas ($N=13$~objects), $0.1 \pm 1.1$~mas ($N=8$~objects) and $-0.9\pm 1.6$~mas ($N=7$ objects), respectively\footnote{In all these calculations we excluded TOK39Aa,Ab, see previous footnote.}, showing that our orbital parallaxes are indeed reliable. It is interesting to note that the rms with respect to Gaia is smaller than that with respect to Hipparcos, which indicates that at least some of the variance on this difference comes from the trigonometric parallaxes themselves, and not from the orbital parallaxes, as the rms is smaller for the better quality Gaia parallaxes.

Several objects of our sample show ambiguity regarding their mass ratio: preliminary results using both the ORBIT routine and our MCMC method gave estimates of $m_{\mbox{B}}/m_{\mbox{A}}$ in regions arbitrarily close to 1. Moreover, in most of these cases swapping the primary and the secondary, i.e., taking observations of RV of the primary as observations of the secondary and vice versa, also led to valid solutions. To deal with this ambiguity, we adapted the methodology described in \citet{Mendezetal2017, Mendet2021} to accept values of $m_{\mbox{B}}/m_{\mbox{A}}$ greater than 1. While both previous studies and the present work rely on a Metropolis-within-Gibbs scheme to generate samples from the posterior distributions, the former rejects samples containing a value of $m_{\mbox{B}}/m_{\mbox{A}}$ outside the interval $(0,1)$. Instead, the method proposed here swaps RV observations whenever $m_{\mbox{B}}/m_{\mbox{A}}$ is greater than 1 and carries out the minimum least-squares estimation of $a$, $\omega$, and $V_{\text{CoM}}$ -conditional to the rest of the parameters- in a manner akin to that explained in \citet{Mendezetal2017}, Appendix~1. This approach produces a shift of $\pm 180^\circ$ in $\omega$ as this parameter explicitly depends on the precedence of binary and secondary which  is corrected in a post-processing step.

\section{HR diagram}\label{sec:hrdiag}

\begin{figure}[ht]
\includegraphics[width=\textwidth, height=12.5cm]{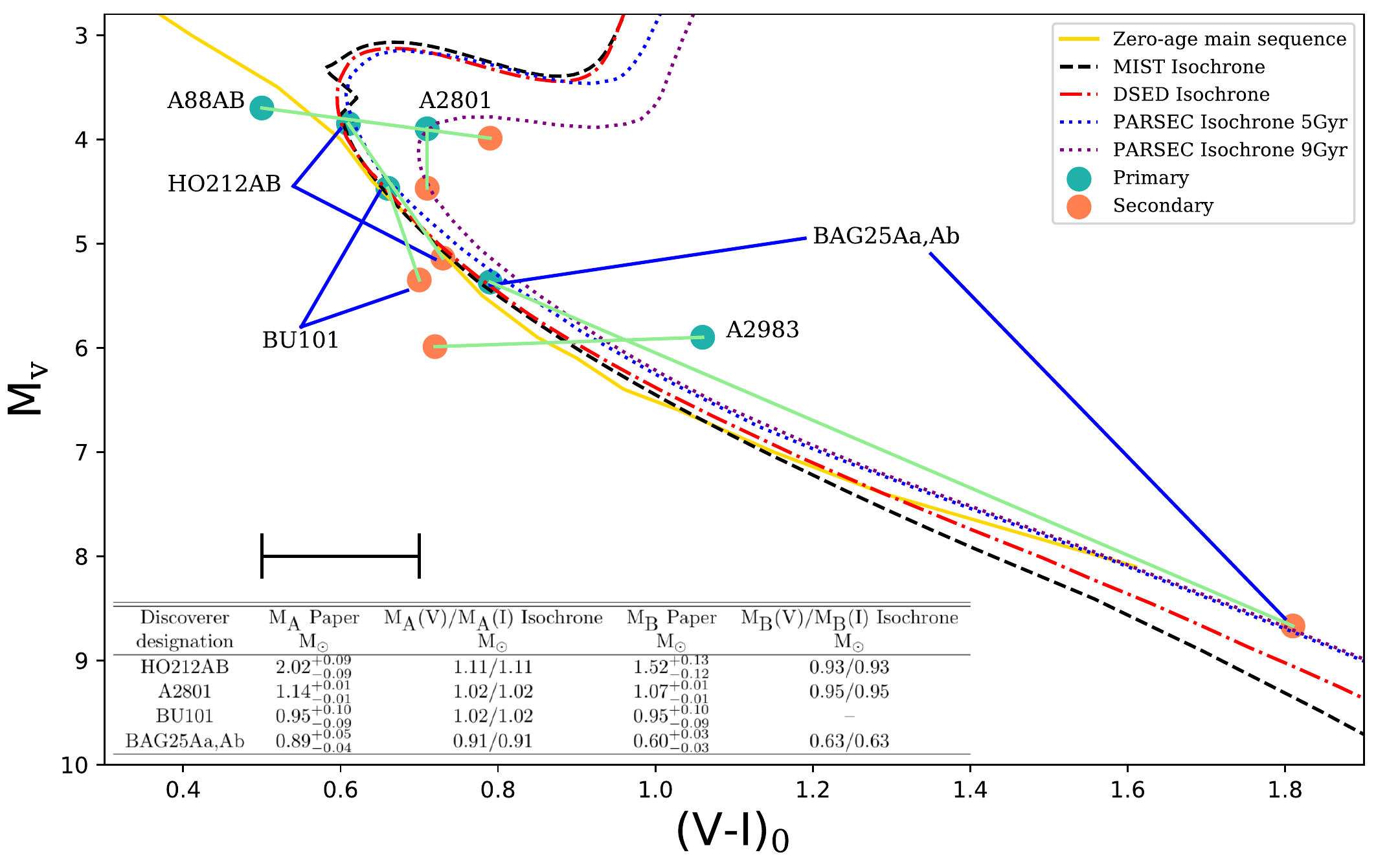}
\caption{HR diagram for our sample of SB2 systems with available photometry. Green dots depict primary components and orange dots the secondaries. Each pair has been linked with a line and the discoverer designation is noted. The bar at (0.6,8.0) shows the estimated error of the photometry, as discussed in Section~\ref{sec:sample}. For reference, we have plotted an empirical Zero-age main sequence, three solar-metallicity ($Z_\odot=0.0152$) 5~Gyr old theoretical isochrones and a 9~Gyr isochrone. In the inserted table we show a comparison of our estimated masses and the mass predicted by the theoretical models. See text for details and comments on individual systems.} \label{fig:hrdiag}
\end{figure}

In Figure~\ref{fig:hrdiag} we present an observational HR diagram for the six visual systems with available $V$ and $(V-I)$ for each component. To obtain the $(V-I)$ colors we used the $V$ magnitudes given in Table~\ref{tab:photom1} and the $I$ magnitudes derived as explained in Section~\ref{sec:comp}. To determine $M_V$, we used the published trigonometric parallaxes given in Table~\ref{tab:plx}. We note that, due to the $\log$ factor, there is no significant difference if we instead use the orbital parallax. Also, at this scale the formal error in absolute magnitude, due to photometric and parallax uncertainties, is negligible 
(of course, this does not consider possible systematic effects or biases on the parallaxes, which could be larger than the formal uncertainties). All these systems lie at a distance less than 65~pc, and hence we did not apply extinction nor reddening correction to the apparent magnitudes and colors.

For reference, in the HR diagram we have superimposed a zero-age main sequence (ZAMS) from  \citet{Sch1982} (gold solid line, kindly provided by G. Carraro\footnote{Personal communication}). In order to asses the current uncertainties in stellar models, we have also superimposed isochrones from the Dartmouth Stellar Evolution Database (DSED\footnote{Available at \url{http://stellar.dartmouth.edu/models/isolf_new.html}}) - as described in \citet{Dotteret2008}, the Padova and Trieste Stellar Evolution Code (PARSEC\footnote{Available at \url{http://stev.oapd.inaf.it/cgi-bin/cmd_3.4}}) - as described in \citet{Bressanet2012} and the MESA Isochrones \& Stellar Tracks code (MIST\footnote{Available at \url{http://waps.cfa.harvard.edu/MIST/interp_isos.html}}) - as described in \citet{Dotteret2016}. While there is an overall good agreement between all these isochrones, the non-zero width of the MS locus for the same age and metallicity shows the impact of using slightly different input physics in the models.

\begin{figure}[ht]
\includegraphics[width=\textwidth, height=12.5cm]{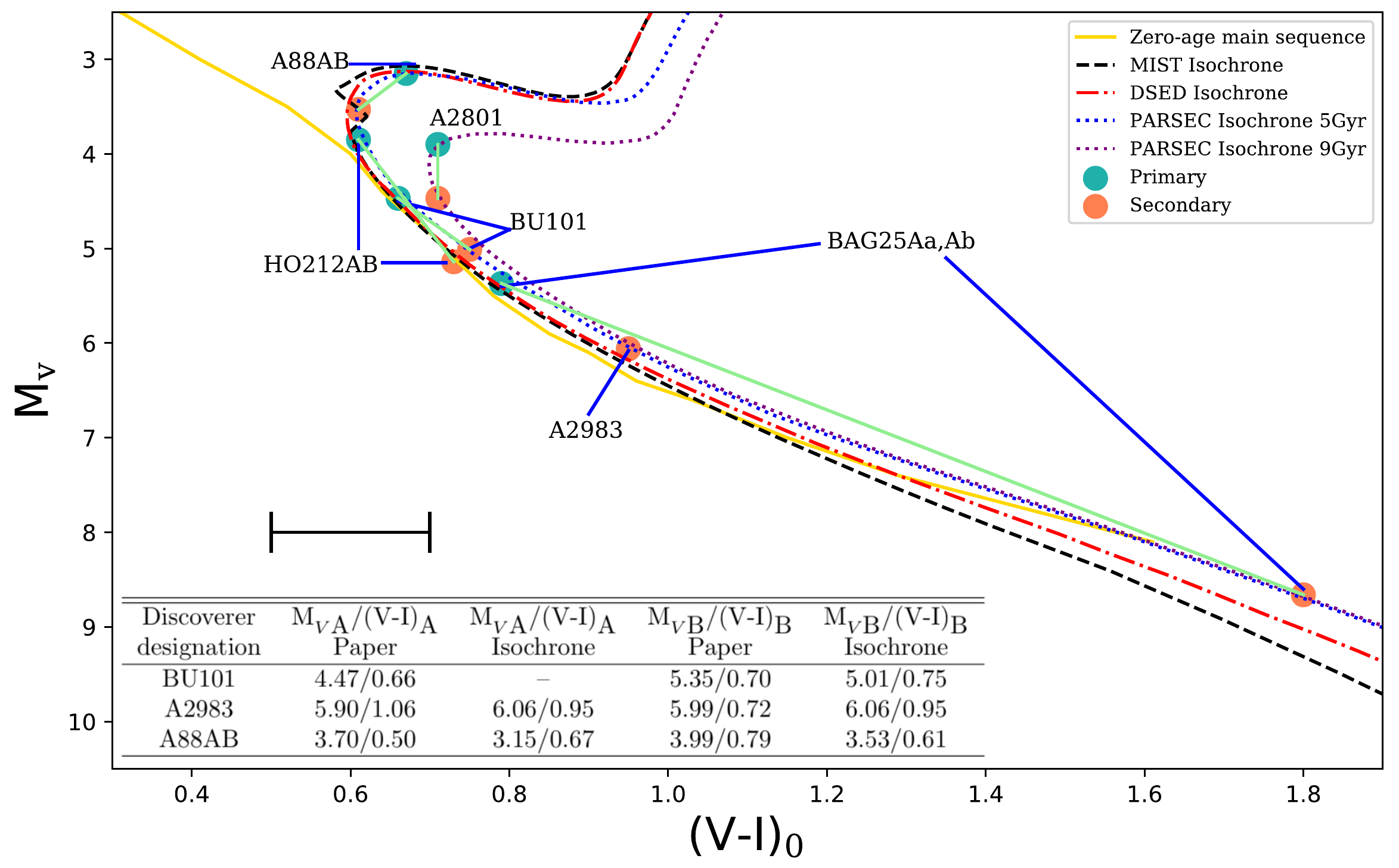}
\caption{Same as Figure~\ref{fig:hrdiag}, for objects with dubious photometry. Having corrected their photometry using the procedure explained in the text we depict their putative location on the HR diagram. The table inserted in the figure gives their measured and "corrected" photometry. \label{fig:hrdiag_2}}
\end{figure}

For the four systems that fall close to the isochrones -implying that their photometry is reliable- we have deduced their mass using the $M_V$ and $M_I$ versus mass relationships obtainable from the isochrones, in order to make a comparison with our dynamical masses. To this end, we used the PARSEC 5~Gyr solar metallicity isochrone, except in the case of A2801 which is better fitted by the PARSEC 9~Gyr isochrone. The comparison is shown in the table inserted in Figure~\ref{fig:hrdiag}. This exercise is obviously not meaningful in the case of the three systems that lie far from the isochrones: WDS 07518$-$1354=BU101, WDS 14492$+$1013=A2983 and WDS 18384$-$0312=A88AB. For these latter we can do a reverse process; that is, starting from the dynamical masses compute the predicted photometry they should have. This approach assumes that the published photometry is erroneous, that our masses are reliable and that the theoretical predictions are accurate. The results of this exercise are shown in Figure~\ref{fig:hrdiag_2}. In the table inserted in this figure we give the resulting "corrected" photometry.  We believe that these discrepancies are not related to metallicity. Even though of the three offending cases there is published metallicity only for A2983 (with $[$Fe/H$4] = -0.03$ according to SIMBAD), we base our conclusion on the fact that the main sequence of the PARSEC theoretical isochrones with lower metallicity, can not simultaneously fit the location on the HR diagram of  the primary and secondary in any of the cases. Further discussions are presented in the next Section, on a case by case basis.

\section{Discussion of individual objects}\label{sec:objnotes}

Based on Figures~\ref{fig:hrdiag} and~\ref{fig:hrdiag_2} in the previous Section, and our orbital fitting results from Section~\ref{sec:orbs}, in what follows we present comments on individual systems.

We note that seven of our systems, identified below, are included in a recent paper by \citet{Picc2020}, who compiled a list of SB2s with published astrometric orbits in order to determine the orbital parallaxes. This group however did not re-compute orbits; using the published orbital elements they derived orbital parallaxes assuming that the (independent) fits to the astrometric orbit and the RV were consistent. This differs substantially from our approach in that we have computed orbital elements in a dynamically self-consistent way by performing a simultaneous fit to all the available data.

\begin{trivlist}
\item {\bf WDS 00352$-$0336=HO212AB:} Ours is the first combined orbit for this SB2 system in which the primary is an SB1 system (not studied here). Since the last published orbit in 2005, we have added 25 new HRCam@SOAR interferometric measurements, the most recent ones on 2018.56 and 2019.95. Our orbital parallax ($42.4 \pm 0.9$~mas) is slightly smaller than that reported by \citet{Picc2020} ($45.2 \pm 1.4$~mas), but they are consistent within $2\sigma$. Our result is however significantly smaller than the Hipparcos parallax at the $5\sigma$ level; not very comfortable considering that both the astrometric orbit and RV curves are quite well sampled and all the orbital elements have small formal uncertainties. We note that a parallax from Gaia is not available yet.

As shown in Figure~\ref{fig:par}, this is actually our most extreme outlier in terms of the difference between the orbital and trigonometric parallax. In SB9 the orbit for this system is currently considered as preliminary, and it was derived keeping the period and eccentricity fixed, which were presumably assumed from the visual orbit. In the present work, these parameters were well determined. Based on the location of both components on the HR diagram, their photometry seems reliable. However, judging from the theoretical isochrones, their inferred masses should be significantly smaller (see table inserted in Figure~\ref{fig:hrdiag}). If we scale down the individual masses reported on Table~\ref{tab:plx} to the Hipparcos parallax instead of our orbital parallax, the individual masses become 1.49 and 1.11~M$_\odot$ respectively, closer to the values inferred from the isochrones but still too large. The spectral type (F7V-F8V) for the primary implies a mass between 1.23 to 1.29~M$_\odot$, while for the secondary (G4V, see Table~\ref{tab:photom1}) it should be 1.06~M$_\odot$ (see Table~18 on \citet{Abuset2020}). These numbers are
still slightly larger than the masses implied by the isochrones, but more in line with the larger Hipparcos parallax than with our orbital parallax. Even if we disregard the mass implied by the spectral type of the primary, we note that in general there is a good correspondence between the different sources of the photometry for this binary, as can be seen in Tables~\ref{tab:photom1} and \ref{tab:photom2}.

We conclude that the implied masses from the isochrone shown on Figure~\ref{fig:hrdiag} are probably reliable, thus at present we have no explanation for the large difference between our dynamical and the isochrone masses.\\

\item {\bf WDS 02128$-$0224=TOK39Aa,Ab:} This object is not listed in Orb6, so ours is the first astrometric orbit and also the first combined orbit. Phase coverage is excellent in RV, but astrometrically it is rather poor;less than 50\% of the orbit has been sampled, which results in particular in a somewhat uncertain inclination of the orbit. See Table~\ref{tab:orbel} and the top panel of Figure~\ref{fig:hip10305pdf}. 
As a consequence, the orbital parallax is not well determined, and the individual component masses exhibit a very large uncertainty. However, we can use the  Gaia parallax as a prior in our solution (see Table~\ref{tab:plx}), which leads to better defined orbital parameters, and well-constrained masses. The main impact of using this prior in the solution is a significant reduction in the uncertainty of the semi-major axis, which varied from $13.98^{+0.75}_{-0.64}$~mas to $14.12^{+0.15}_{-0.15}$~mas. Being this a first orbit, we show the fits to this system in Figure~\ref{fig:hip10305}, while in Figure~\ref{fig:hip10305pdf} we show the PDFs without (upper panel) and with (lower panel) the use of a parallax prior. Examining the values on Table~\ref{tab:plx} and the PDFs, it is interesting to note that despite the fact that there is a significant reduction in the interquartile ranges when using a parallax prior, the ML and mid-quartiles are not that different between these two solutions, i.e., the best estimates seem to be somewhat resilient to uncertainties in the orbital fitting. Nevertheless, we expect to improve the astrometric orbit in the next few years; the most recent epochs are from our programs with HRCam@SOAR in 2020.82 and 2020.92, and with ZORRO@GS\footnote{For a description of the ZORRO instrument and its reduction pipeline, please see \citet{Howellet2011}, \citet{Horchet2011}, and \citet{Scottet2018}.} in 2020.83. Unfortunately, all our latest observations cover the same sector of the orbit (see Figure~\ref{fig:hip10305}). As a final note, the F8V spectral type for the primary implies a mass of $1.23 \pm 0.05$~M$_\odot$ (\citet{Abuset2020}, Table~18), within 1$\sigma$ of our dynamical mass (with prior) as shown in Table~\ref{tab:plx}.\\

\begin{figure}[h]
\centering
\includegraphics[width=0.465\textwidth]{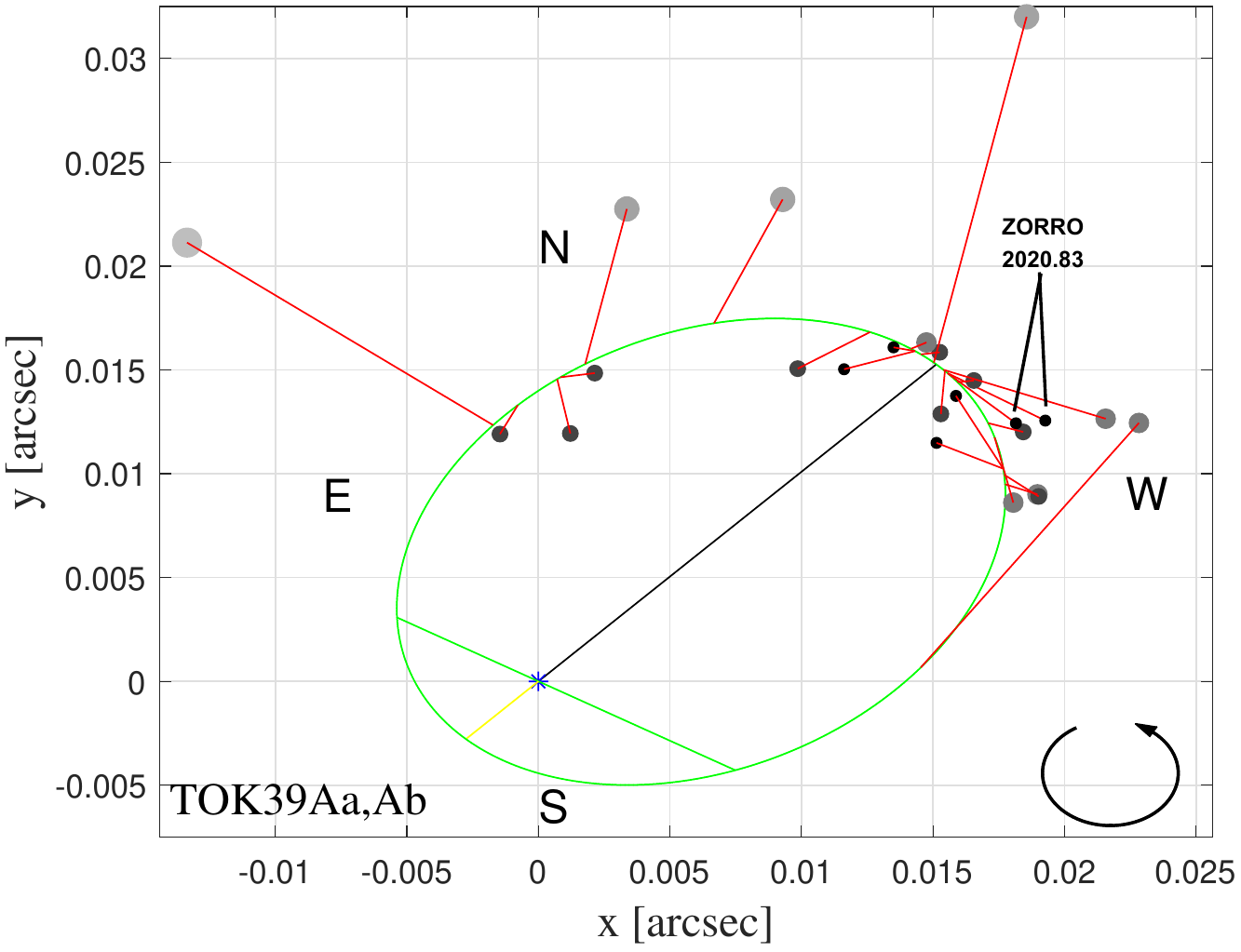}
\includegraphics[width=0.45\textwidth]{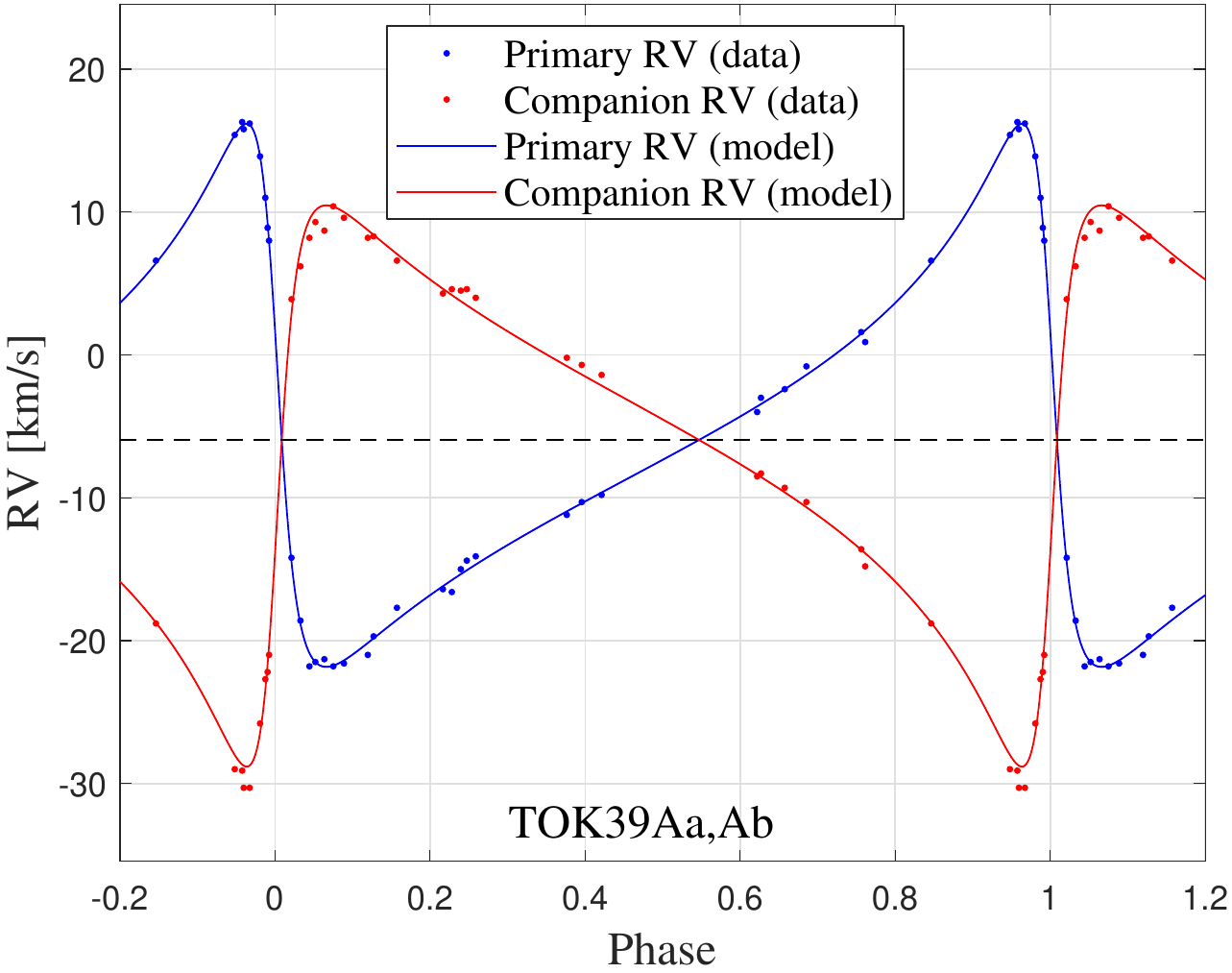}
\caption{Maximum likelihood orbit for TOK39Aa,Ab. Symbols as in Figure~\ref{fig:examples}. Shown is the solution without a prior. As explained in the text, if a parallax prior is used the resulting plot is quite similar. The currently incomplete orbital coverage justifies the use of a high-precision Gaia parallax prior, which improved significantly the precision of our estimate of the individual component masses (see Figure~\ref{fig:hip10305pdf}).
\label{fig:hip10305}}
\end{figure}

\begin{figure}[h]
\centering
\boxedfig{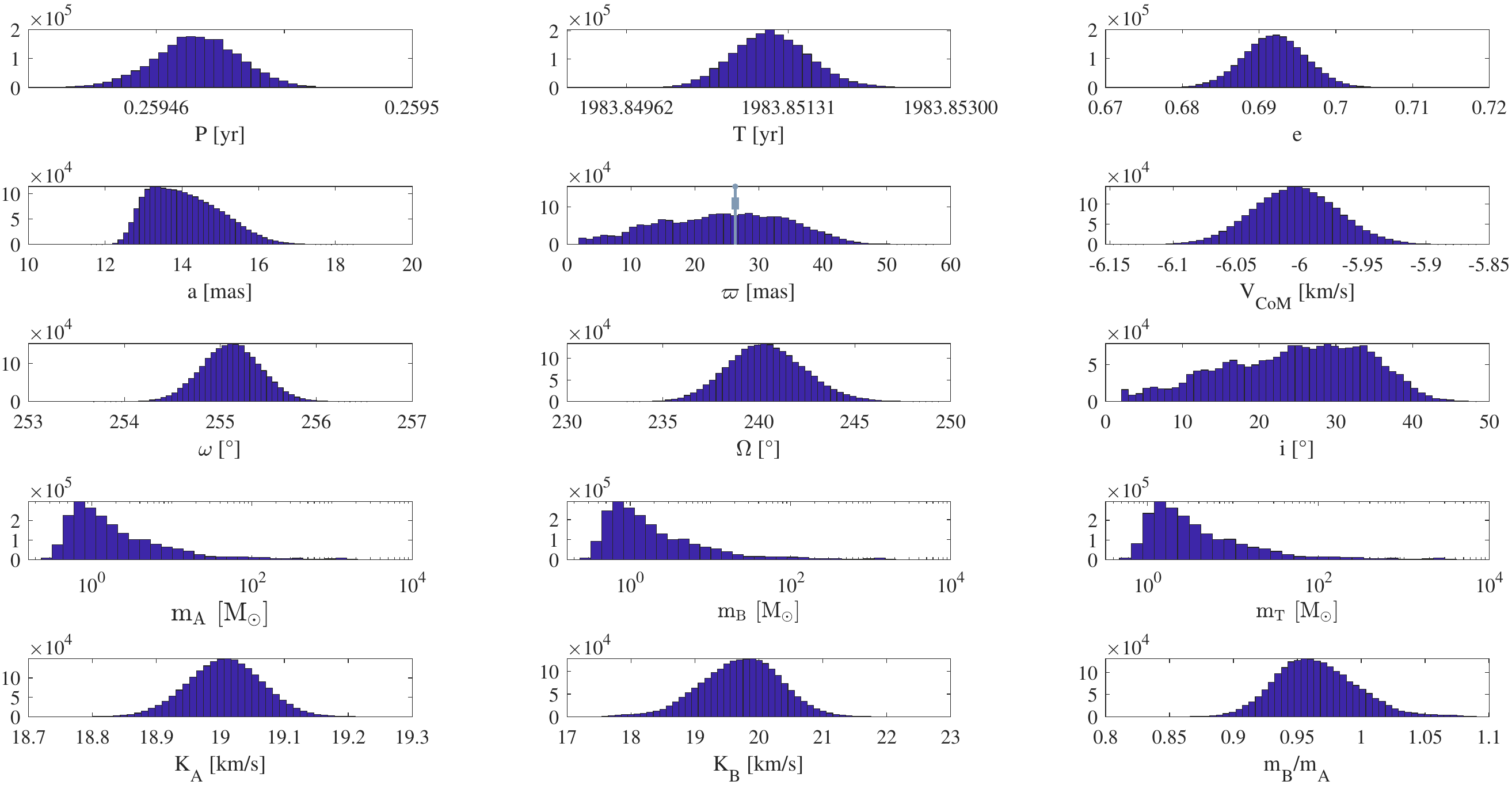}{1.0\textwidth}{\textbf{TOK39Aa,Ab - No prior}}\\
\boxedfig{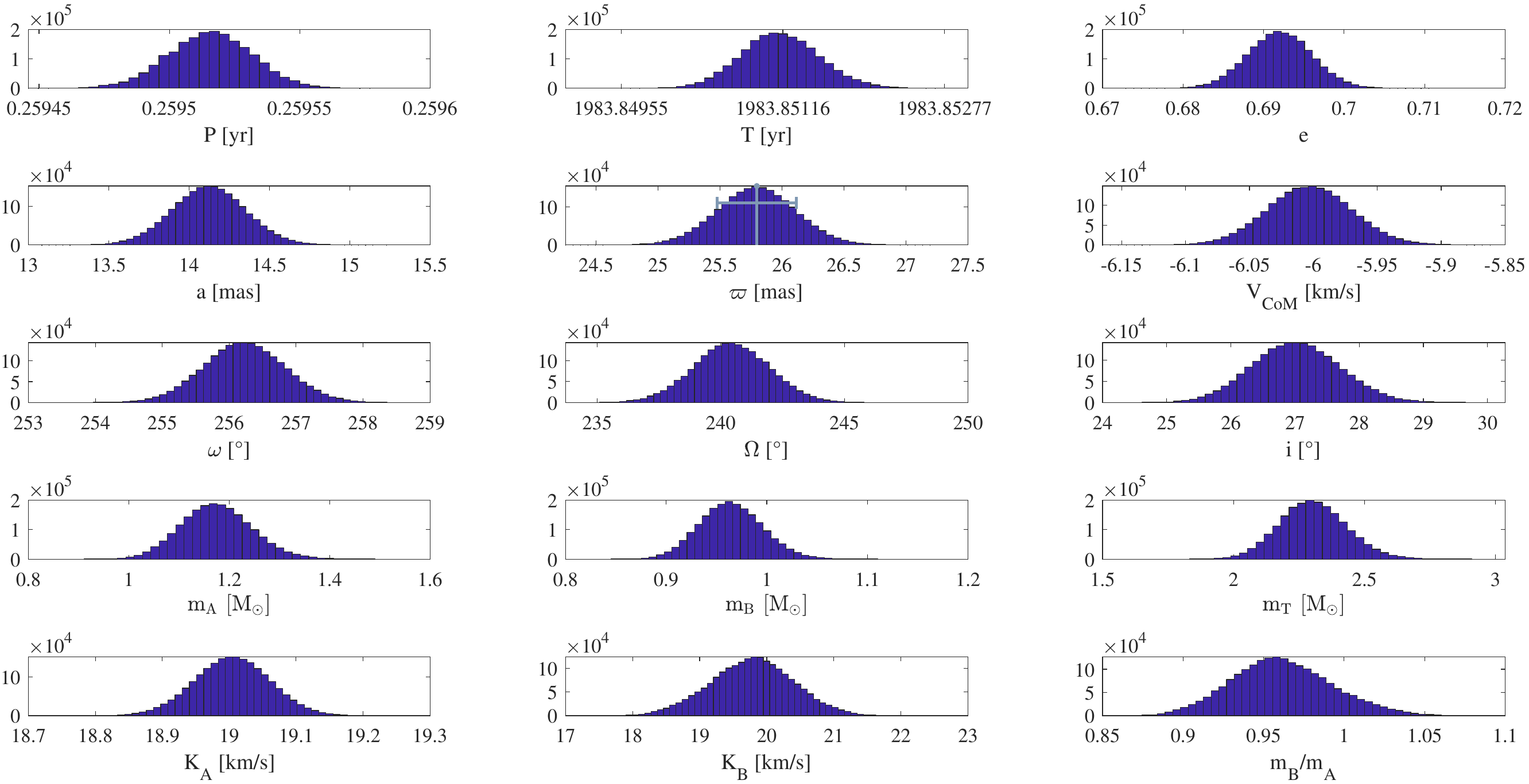}{1.0\textwidth}{\textbf{TOK39Aa,Ab - Prior}}\\
\caption{Posterior distribution functions for TOK39Aa,Ab. The top panel shows the solutions obtained without a parallax prior, and the lower panel those obtained with a parallax prior. Note that in the upper panel the mass scale is logarithmic, while in the lower panel it is linear. Even in the no-prior scenario the inter-quartile range and the ML value for the parallax is commensurate with the Gaia eDR3 value.\label{fig:hip10305pdf}}
\end{figure}

\item {\bf WDS 04107$-$0452=A2801:} The latest astrometric orbit available for this system is from \citet{Tok2017}. We have added two new measurements made on 2016.96 and 2018.97 with HRCam@SOAR, and obtained an orbital parallax of $16.58 \pm 0.12$~mas. This object is also included on Piccotti's study, who obtained an orbital parallax of $17.12 \pm 0.24$~mas, while a previous study from \citet{Docet2017} gives $16.18 \pm 0.23$~mas. This latter value is more in line with our result. 

Our derived individual component masses are quite consistent with those from \citet{Docet2017} and Piccotti, but somewhat larger than those implied by the isochrones (see table inserted in Figure~\ref{fig:hrdiag}). If we scale our masses to the Gaia parallax instead of our orbital parallax, the individual component masses turn out to be 1.10 and 0.98~M$_\odot$ respectively, quite close to those from the isochrones. However, the spectral type of the primary (G0IV to G5IV) implies a mass between 1.26 to 1.20~M$_\odot$ (see Table~19 on \citet{Abuset2020}), closer to the masses derived from our orbital parallax. The photometry suggests that the primary is leaving the main-sequence; in our HR diagrams the isochrone that best fits both components is the PARSEC 9~Gyr (see Figure~\ref{fig:hrdiag}). This is consistent with the luminosity class IV given for the primary in SIMBAD (see Table~\ref{tab:photom1}).\\

\item {\bf WDS 04184$+$2135=MCA14Aa,Ab:} This is a re-analysis of an orbit already studied by \citet{Torres1997}, who derived a combined spectroscopic and astrometric solution, yielding an orbital parallax of $17.92 \pm 0.58$~mas, and masses of $1.80 \pm 0.13$~M$_\odot$ and $1.46 \pm 0.18$~M$_\odot$ for primary and secondary, respectively. More recently, \citet{Pour2000} has also obtained a combined solution for this resolved SB2, yielding orbital parallaxes and component masses similar to those derived by Torres and collaborators. After these studies, five new astrometric observations of this system have been secured between 1997.14 and 2005.86, using 4m facilities with adaptive optics and speckle imaging, and we included them in our re-analysis. Our derived orbital parallax is sightly larger, at $18.16 \pm 0.39$~mas, but our individual masses are basically the same as those from \citet{Torres1997}. We note that the Gaia eDR3 trigonometric parallax is larger than ours by almost $3\sigma$ (see Figure~\ref{fig:par}). \citet{Picc2020} give an orbital parallax of $17.55\pm 0.59$~mas, even more discrepant with the eDR3 Gaia value, albeit their derived masses are not so discrepant from ours: $1.87 \pm 0.58$ and $1.52 \pm 0.19$~M$_\odot$. Our formal errors are however significantly smaller (see Table~\ref{tab:plx}). Given the spectral type of the primary (F0V) \citet{Abuset2020} predict for it a mass of $1.64 \pm 0.05$~M$_\odot$, which differs by less than 1$\sigma$ of our dynamical mass.\\

\item {\bf WDS 07518$-$1354=BU101:} We have included this well-studied equal-mass binary as a benchmark, to compare literature values with our results. The result reported in SB9 comes from a combined astrometric and spectroscopic orbit by \citet{Pour2000}. The visual orbit has been subsequently revised by \citet{Tok2012b}, using newer astrometric observations made with HRCam@SOAR. Our solution incorporates more recent measurements made with the same sert-up\footnote{Actually, this binary is observed frequently as an "astrometric standard" because it is used to calibrate the plate-scale and orientation of HRCam - see Section~\ref{sec:orbs}.}, being 2019.95 our last epoch. Our orbital parameters are in agreement with those from Tokovinin, albeit with smaller formal uncertainties on account of the incorporation of new data. Given its photometry, the mass of the primary is in very good agreement with the prediction from the isochrones (see table inserted in Figure~\ref{fig:hrdiag}), but the secondary is off, which casts some doubts on the photometry of this latter. In Figure~\ref{fig:hrdiag_2}, which includes the photometry in the inserted table, we show the location that the secondary should have had given its empirical mass and the isochrones. The difference between the measured and expected magnitudes in $V$ is 0.34~mag, which is quite large, albeit in color the difference is smaller, 0.05~mag. Note also that the $V$-band photometry for this target does not exhibit such a large variance according to Tables~\ref{tab:photom1} and ~\ref{tab:photom2}. Our individual masses seem more consistent with a G7V to G8V spectral type (according to \citet{Abuset2020}, their Table 18), rather than with the earlier types given in WDS (G0V) and SIMBAD (G1V) which imply larger masses. However, given our 10\% mass uncertainty, the earlier types are consistent within $1\sigma$ of our mass interquartile range (see Table~\ref{tab:plx}). We finally note that \citet{Picc2020} reports an orbital parallax of $64.4 \pm 2.4$~mas, which is within $1.5\sigma$ of our value of $60.2 \pm 1.5$~mas; and the same happens with the individual component masses. Perhaps the most intriguing aspect of this binary is that, despite the fact that both components have the sames mass (within the errors), their photometry seems to indicate different locations for them on the HR diagram which should not be the case if they are coeval: Note, e.g., the good correspondence between $\Delta V$ and the $\Delta y$ value given in Table~\ref{tab:photom1}, which shows that they do not seem to have the same luminosity.\\

\item {\bf WDS 11560$+$3520=CHR258:} Ours is the first combined orbit for the external pair of this triple hierarchical system. The formal uncertainty of the orbital parallax (1.9~mas) is significantly worse than the uncertainties of the trigonometric parallaxes from Hipparcos (0.58~mas) and Gaia eDR3 (0.37~mas). This is most likely due to the poor orbital coverage: only three astrometric points are available - albeit they are well distributed in the orbit. An alternative solution, using a fixed parallax at the eDR3 value, leads to larger individual component masses of 1.84 and 1.53~M$_\odot$, in accordance with the smaller system's parallax. The spectral type (F5) implies however a mass of $1.39 \pm 0.05$~M$_\odot$ (\citet{Abuset2020}), in agreement with our value of $1.22^{+0.45}_{-0.26}$~M$_\odot$, thus somewhat validating our larger orbital parallax. This is an interesting system that deserves further astrometric observations of the external pair for a better orbital coverage. The phase coverage on the RV curve is already quite good.\\

\item {\bf WDS 14492$+$1013=A2983:} The latest visual orbit included in Orb6 is that from \citet{2018IAUDS}, but more recently \citet{AlTawalet2021} has revised this solution, and obtained a system's mass of $1.61 \pm 0.26$~M$_\odot$, and an orbital parallax of $21.81 \pm 0.8$~mas. These results lie within $1\sigma$ of our derived values, albeit our combined solution yields much smaller formal uncertainties, due in part to our combined solution and also to the addition of three new HRCam+SOAR measurements in 2018.16, 2019.14, and 2019.54. The mass implied by the spectral type of the primary (K2V) is $0.80 \pm 0.03$~M$_\odot$ (\citet{Abuset2020}), in very good agreement with our dynamical mass. \citet{Griffin2015} reports \textbf{$m_{\mbox{A}} \sin^3 i=  0.308 \pm 0.012$ and $m_{\mbox{B}} \sin^3 i = 0.306 \pm 0.011$~M$_{\odot}$} for this object, which for our value of the inclination (see Table~\ref{tab:orbel}) implies masses of 0.898 and 0.892~M$_\odot$, consistent at the 1$\sigma$ level with our determination shown in Table~\ref{tab:plx}. 
As can be seen in Figure~\ref{fig:hrdiag}, the measured photometry seems however at odds with its location on the HR diagram, in particular regarding the large $\Delta I = 0.43 \pm 0.12$~mag given in Table~\ref{tab:photom1}. In Figure~\ref{fig:hrdiag_2} we show the location it should have on the HR diagram on account of its mass; because it is an equal-mass binary, the primary and secondary are located on the same point of the isochrone.\\

\item {\bf WDS 15282$-$0921=BAG25Aa,Ab:} This is the first combined orbit for this highly eccentric SB2 inner pair of a triple system. \citet{Picc2020} gives masses of $0.84 \pm 0.30$ and $0.58 \pm 90.21$~M$_\odot$ and an orbital parallax of $52.7 \pm 4.7$~mas. Our orbital parallax is smaller and more in line with the Hipparcos parallax (no Gaia parallax for this object yet), and with formal errors a factor of two smaller in orbital parallax, and a factor of six improvement in the individual masses. This is mostly due to the fact that, since the last published orbit in Orb6 from \citet{Tok2016}, our survey has added seven high-precision observations from HRCam@SOAR, the latest being on 2019.14.  The photometry seems reliable; all measurements on Tables~\ref{tab:photom1} and \ref{tab:photom2} agree within the uncertainties As shown in the table inserted on Figure~\ref{fig:hrdiag}, our orbital masses agree very well with the theoretical masses for both the primary and secondary. Based on the spectral type (G9V) of the primary, \citet{Abuset2020} predict a mass of $0.93 \pm 0.04$~M$_\odot$, consistent with our reported value is of $0.89^{+0.05}_{-0.04}$~M$_\odot$. \\

\item {\bf WDS 16584$+$3943=COU1289:} Ours is the first combined orbit for this system. \citet{Picc2020} obtained an orbital parallax of $9.00 \pm 0.30$~mas, slightly smaller than ours (but within 1$\sigma$) , and more in line with the Gaia eDR3 value. Their individual masses are also within 1$\sigma$ with ours, but our formal errors are much smaller. No new observations were incorporated to our solution; just the data used in the original visual orbit by \citet{2013IAUDS}, and the spectroscopic orbit from \citet{Griff2003}. The differences in uncertainty with respect to the Piccotti result are likely a consequence of the better performance of the self-consistent simultaneous combined fit. Given the small errors of both the Gaia parallax and our orbital parallax, the difference of 0.36~mas seems uncomfortably large (see Figure~\ref{fig:par}). Note that the RV curve covers the range 0.9 to 1.1 in phase -with no data at intermediate phases- while almost the opposite happens in the case of the astrometric curve; which may be the culprit for a somewhat uncertain orbital parallax. If we assume the primary is a G0V, its $M_V$ results to be $+4.40$ (\citet{Abuset2020}, their Table~18), which for the Gaia parallax (see Table~\ref{tab:plx}) implies a primary $V=9.64$. If we assume a nearly equal-mass system, consistent with our results given in Table~\ref{tab:plx}, and with
$\Delta V = 0$ reported by WDS, then the system's magnitude would be $V=8.88$, which is significantly fainter than the magnitude predicted by WDS (7.65~mag) and also fainter than the Hipparcos measurement (8.09~mag). Incidentally, ASAS-SN reports $V=8.52 \pm 0.094$, but this value is uncertain due to the bright limit at $V=10$ of this survey. We have no explanation for this discrepancy on the photometry.\\

\item {\bf WDS 18384$-$0312=A88AB:} First combined orbit. Very tight fit with a good orbital and phase coverage. Our results are in agreement with those from \citet{Malkovet2012} who obtain a dynamical masses for the pair of $2.42 \pm 0.32$~M$_{\odot}$. On the other hand, \citet{Griffin2013} obtained from the RV alone $m_{\mbox{A}} \sin^3 i=  0.690 \pm 0.013$ and $m_{\mbox{B}} \sin^3 i = 0.660 \pm 0.012$~M$_{\odot}$, which for our value of the inclination (see Table~\ref{tab:orbel}) imply masses of 1.75 and 1.15~M$_\odot$. The mass for the primary is significantly larger than ours (1.20~M$_\odot$), whereas the mass of the secondary is totally consistent with our result. Given the spectral type of the primary, F8V-F9V, the implied mass (\citet{Abuset2020}) is 1.23-1.20~M$_{\odot}$, which coincides well with our value.

We note that since the last visual orbit published for this binary in 2013, eight new interferometric observations have been made; two on 2014.76 by \citet{Horchetal2015} with the 4.3m Discovery Channel Telescope and six by us with  HRCam@SOAR program, between 2015.50 and 2019.61.

As result, both the RV curve and the astrometric orbit are now very well sampled, and our orbital parameters and derived masses are tight. There is no Gaia parallax yet for this target, but our orbital parallax coincides within 1$\sigma$ with the Hipparcos parallax. As it was the case for A2983, the measured photometry seems at odds with its location on the HR diagram (see Figure~\ref{fig:hrdiag}). As before, in Figure~\ref{fig:hrdiag_2} we show the location it should have on the HR diagram given its mass, together with the photometry in the inserted table. If we assume a typical age for disk stars of 5~Gyr, this location implies that it may be a slightly evolved system.\\

\item {\bf WDS 20102$+$4357=STT400:} A first combined orbit was published by \citet{Pour2000}. In our solution we have included twelve new interferometric observations carried out in the period 2000.20 to 2013.83 by various authors, which improved the orbital coverage. We note that the orbit is not complete yet, because of the period which is nearly 85~years. At this point, the combined solution is limited mostly by the small phase coverage in the RV curve. \citet{Picc2020} gives $17.8 \pm 1.3$~mas for the orbital parallax, and masses of $1.24 \pm 0.38$ and $0.96 \pm 0.26$~M$_\odot$; similar to our values. There are spectral types for both the primary (G3V to G4V) and secondary (G8V) with indicate mass of 1.07-1.06$\pm 0.04$~M$_\odot$ (\citet{Abuset2020}), that compare well with our values within our rather large mass uncertainties of 0.1 to 0.2~M$_\odot$ for this system.\\

\item {\bf WDS 20205$+$4351=IOT2Aa,Ab:} We selected this interesting target because of its small, and hence uncertain trigonometric parallax (even for Gaia). On the other hand, in principle, our orbital parallaxes are distance-independent. The O-type + WR pair has been studied by \citet{Monnier2011}, who derived masses of $35.9 \pm 1.3$ and $14.9 \pm 0.5$~M$_\odot$, for a distance of $1.76 \pm 0.03$~kpc; i.e., a parallax of $0.57\pm 0.01$~mas. The latest Gaia eDR3 parallax is $0.538 \pm 0.069$~kpc (see Table~\ref{tab:plx}), is consistent within 1.5$\sigma$ with the \citet{Monnier2011} value. Our estimated orbital parallax is larger than the previous values, implying smaller component masses of $20.5 \pm 3.2$ and $8.4 \pm 1.6$~M$_\odot$. The precision on our orbital parallax is however a factor of two worse than that from Gaia (see Figure~\ref{fig:par}), probably due to the combination of the rather poor orbital coverage with noisy RV of the early-type stars involved\footnote{See the RV fits on \url{http://www.das.uchile.cl/~rmendez/B_Research/JAA-RAM-SB2/}}.
Scaling our parallax to that of \citet{Monnier2011} would imply masses of 36.5 and 15.0~M$_\odot$, more in-line with their results. Unfortunately, our attempt to use this system as a test case of the orbital parallaxes is not conclusive, not due to a fundamental limitation of our methodology, but rather due to the limited coverage and the quality of the observational data.\\

\item {\bf WDS 20527$+$4607=A750:} Ours is the first combined orbit. The visual orbit only encompasses about 50\% of the orbit, but the combined solution looks solid, with small uncertainties. Our orbital parallax coincides within $1\sigma$ with the Gaia eDR3 result, while the individual component masses have a 4\% uncertainty. We have obtained $1.08 \pm 0.05$~M$_\odot$ for the mas of the primary, which coincides reasonably well with that predicted by \citet{Abuset2020} from the G8V spectral type of the primary ($0.96 \pm 0.04$~M$_\odot$).\\

\item {\bf WDS 23485$+$2539=DSG8:} Ours is the first fully self-consistent orbit. \citet{Horch2019} provided orbital elements by fitting the RV curve independently from the visual orbit obtained using Tokovinin's ORBIT code (see Table~5 on \citet{Horch2019}). They obtained an orbital parallax $14.3 \pm 0.4$~mas, similar to our value; and likewise with the masses that have very small uncertainties. At $\sim$F5V, the WDS spectral types of the primary and secondary, imply a large mass of $1.35$M$_\odot$, (\citet{Abuset2020}). SIMBAD gives a spectral type of F2IV-V indicating masses even larger. This discrepancy is probably due to the low-metallicity of the system ([Fe/H]=$-$0.46) which makes the spectral type of the system seem earlier than it really is.\\

\end{trivlist}

\section{Conclusions} \label{sec:concl}

We have done a thorough search of the Sixth Catalog of Orbits of Visual Binary Stars (Orb6, \citet{WDSCat2001}) and the 9th Catalog of Spectroscopic Binary Orbits (SB9, \citet{Pouret2004}) looking for double-line spectroscopic binaries lacking a published combined visual and spectroscopic orbit. We found eight systems which met this condition: WDS00352$-$0336=HO212AB, WDS02128$-$0224=TOK39Aa,Ab, WDS11560$+$3520=CHR258, WDS15282$-$0921=BAG25Aa,Ab, WDS16584$+$3943=COU1289, WDS18384$-$0312=A88AB, WDS20527$+$4607=A750, and WDS23485$+$2539=DSG8. One of these pairs, TOK39Aa,Ab, also did not have a visual orbit.

Using a MCMC code developed by our group, we carried out dynamically self-consistent simultaneous fits to the data, obtaining orbital elements, individual component masses and orbital parallaxes. A comparison of our orbital parallaxes with trigonometric parallaxes from Hipparcos and Gaia, shows in general a good agreement.

We also computed joint solutions for six comparison binaries: WDS04107-0452=A2801, WDS 04184+2135=MCA14Aa,Ab, WDS07518$-$1354=BU101, WDS14492$+$1013=A2983, WDS20102$+$4357= STT400, and WDS20205$+$4351=IOT2Aa,Ab.
Even in these cases we could improve the previous orbits by adding recent data from our Speckle survey
of binaries being carried out with HRCam@SOAR and ZORRO@GS. 

The mass ratios could be determined in the best cases with less than 1\% uncertainty, while the uncertainty on the mass sum is about 1\%. The formal uncertainty of the best individual component masses that we could determine is $0.01 M_\odot$. We have placed those objects that have individual component photometry on an HR diagram, to compare their location in relation to various theoretical isochrones and an empirical ZAMS. We also provide a detailed discussion of our results on an object-by-object basis.

\section{Acknowledgments}

We are very grateful to the referee for his/her careful reading of our manuscript and for his/her many detailed suggestions which have improved the readability of the text.

JAA, RAM and EC acknowledge support from ANID/FONDECYT Grant Nr. 1190038.

We are grateful to Andrei Tokovinin (CTIO/NOIRLab) for his help and support on the use of HRCam at SOAR, to Ricardo Salinas (GS/ZORRO instrument scientist), Elise Furlan (Scientist, NASA Exoplanet Science Institute Caltech/IPAC), and Steve Howell (scientist Space Science and Astrobiology Division, NASA Ames Research Center) for their help and support on the use of ZORRO at Gemini South. We also acknowledge all the support personnel at CTIO and GS for their commitment to operations during these difficult COVID times.

This research has made use of the Washington Double Star Catalog maintained at the U.S. Naval Observatory and of the SIMBAD database, operated at CDS, Strasbourg, France. This research was made possible through the use of the AAVSO Photometric All-Sky Survey (APASS), funded by the Robert Martin Ayers Sciences Fund and NSF AST-1412587. This work has made use of data from the European Space Agency (ESA) mission Gaia (\url{https://www.cosmos.esa.int/gaia}), processed by the Gaia Data Processing and Analysis Consortium (DPAC, \url{https://www.cosmos.esa.int/web/gaia/dpac/consortium}). Funding for the DPAC has been provided by national institutions, in particular the institutions participating in the Gaia Multilateral Agreement. We are very grateful for the continuous support of the Chilean National Time Allocation Committee under programs CN2018A-1, CN2019A-2, CN2019B-13, CN2020A-19, CN2020B-10 and CN2021B-17 for SOAR, and the Gemini Time Allocation Committee under program ID GS-2019A-Q-110, GS-2019A-Q-311, GS-2019B-Q-116, GS-2019B-Q-223, GS-2020A-Q-116, and GS-2020B-Q-142.

Some of the Observations in this paper made use of the High-Resolution Imaging instrument ZORRO. ZORRO was funded by the NASA Exoplanet Exploration Program and built at the NASA Ames Research Center by Steve B. Howell, Nic Scott, Elliott P. Horch, and Emmett Quigley. ZORRO is mounted on the Gemini South telescope of the international Gemini Observatory, a program of NSF’s NOIRLab, which is managed by the Association of Universities for Research in Astronomy (AURA) under a cooperative agreement with the National Science Foundation on behalf of the Gemini Observatory partnership: the National Science Foundation (United States), National Research Council (Canada), Agencia Nacional de Investigaci\'on y Desarrollo (Chile), Ministerio de Ciencia, Tecnolog\'{i}a e Innovaci\'on (Argentina), Minist\'erio da Ci\^{e}ncia, Tecnologia, Inova\c{c}\~{a}es e Comunica\c{c}\~{o}es (Brazil), and Korea Astronomy and Space Science Institute (Republic of Korea).

We humbly dedicate this work to the memory of Dr. Dimitri Pourbaix, a lead researcher in the field of binary-star research, and principal author and keeper of the SB9 catalogue, without which this work would have not been possible. His sudden death, on November 14th 2021, is a great loss to our field, he will be sorely missed by all of us.

\bibliography{Wamba_1}{}
\bibliographystyle{aasjournal}



\end{document}